\documentclass[preprint,superscriptaddress,longbibliography,floatfix]{revtex4-1}
\usepackage{amsmath}
\usepackage{graphicx}
\usepackage{epstopdf}
\usepackage{parskip}
\usepackage[usenames,dvipsnames,svgnames,table]{xcolor}
\usepackage[utf8]{inputenc}
\usepackage{hyperref}
\usepackage{braket}
\usepackage{pbox}
\usepackage{footmisc}

\usepackage{natbib}

\usepackage[english]{babel}

\newcommand{\abinitio}{ab initio}

\begin{document}

	\title{Competing Coulomb and electron-phonon interactions in NbS$_2$}

	\author{Erik G. C. P. van Loon}
	\affiliation{
		Institute for Molecules and Materials,
		Radboud University, 
		NL-6525 AJ Nijmegen, The Netherlands
	}
	\email{evloon@science.ru.nl}
	\author{Malte R\"osner}
	\affiliation{
		Department of Physics and Astronomy, 
		University of Southern California, 
		Los Angeles, 
		CA 90089-0484, USA}
	\affiliation{
		Institut f{\"u}r Theoretische Physik, 
		Universit{\"a}t Bremen, 
		Otto-Hahn-Allee 1, 
		D-28359 Bremen, Germany
	}
	\affiliation{
		Bremen Center for Computational Materials Science, 
		Universit{\"a}t Bremen, 
		Am Fallturm 1a, 
		D-28359 Bremen, Germany
	}
	\author{Gunnar Sch\"onhoff}
	\affiliation{
		Institut f{\"u}r Theoretische Physik, 
		Universit{\"a}t Bremen, 
		Otto-Hahn-Allee 1, 
		D-28359 Bremen, Germany
	}
	\affiliation{
		Bremen Center for Computational Materials Science, 
		Universit{\"a}t Bremen, 
		Am Fallturm 1a, 
		D-28359 Bremen, Germany
	}
	\author{Mikhail I. Katsnelson}
	\affiliation{
		Institute for Molecules and Materials,
		Radboud University, 
		NL-6525 AJ Nijmegen, The Netherlands
	}
	\author{Tim O. Wehling}
	\affiliation{
		Institut f{\"u}r Theoretische Physik, 
		Universit{\"a}t Bremen, 
		Otto-Hahn-Allee 1, 
		D-28359 Bremen, Germany
	}
	\affiliation{
		Bremen Center for Computational Materials Science, 
		Universit{\"a}t Bremen, 
		Am Fallturm 1a, 
		D-28359 Bremen, Germany
	}
	
	\date{\today}
	
	\begin{abstract}		
The interplay of Coulomb and electron-phonon interactions with thermal and quantum fluctuations facilitates rich phase diagrams in two-dimensional electron systems. Layered transition metal dichalcogenides hosting charge, excitonic, spin and superconducting order form an epitomic material class in this respect. Theoretical studies of materials like NbS$_2$ have focused on the electron-phonon coupling whereas the Coulomb interaction, particularly strong in the monolayer limit, remained essentially untouched. Here, we analyze the interplay of short- and long-range Coulomb as well as electron-phonon interactions in NbS$_2$ monolayers. 
The combination of these interactions causes electronic correlations that are fundamentally different to what would be expected from the interaction terms separately.
The fully interacting electronic spectral function resembles the non-interacting band structure but with appreciable broadening. 
An unexpected coexistence of strong charge and spin fluctuations puts NbS$_2$ close to spin and charge order, suggesting monolayer NbS$_2$ as a platform for atomic scale engineering of electronic quantum phases.	
	\end{abstract}
	
	\maketitle

	\section{Introduction}
	
\begin{figure}[hbt]
		\centering
		\includegraphics[width=\columnwidth]{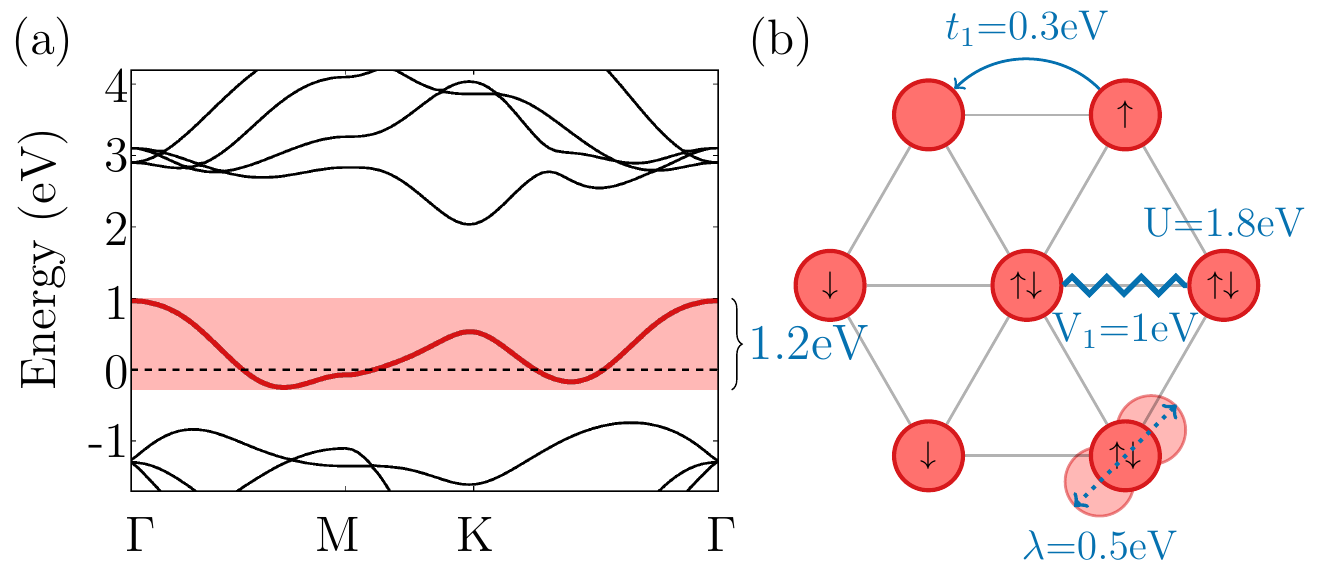}
		\caption{Relevant energies in monolayer NbS$_2$. (a) Band structure. The metallic half-filled band of bandwidth $1.2\,$eV is colored in red. (b) We characterize the electronic properties of NbS$_2$ by the competition between kinetic energy $t$, local electronic repulsion $U$, long-range Coulomb interaction $V$ and on-site electron-phonon interaction $\lambda$. For the kinetic energy and the long-range Coulomb interaction, only the nearest-neighbor terms are visualized, while all terms are taken into account in the numerical simulations.}
		\label{fig:introduction}
	\end{figure}

	Layered materials host in many cases pronounced electronic interaction phenomena ranging from eV-scale excitonic binding energies in semiconductors to charge, spin, and superconducting order in metallic systems. Characteristic energy scales and transition temperatures associated with these interaction phenomena change often remarkably when approaching the limit of atomically thin materials~\cite{yu_gate-tunable_2015, xi_strongly_2015, calandra_effect_2009}. Two generic contributions to these material-thickness dependencies are quantum-confinement~\cite{novoselov_twodimensional_2005, RevModPhys.81.109, geim_van_2013} and enhanced local and long-ranged Coulomb interactions in monolayer thin materials~\cite{Animalu_nonlocal_1964, Keldysh_coulomb_1979, PhysRevB.84.085406, Kirsten_Dielectric_2015, rosner_wannier_2015, PhysRevB.93.235435}. In addition, many layered materials feature sizable electron-phonon coupling~\cite{nagamatsu_superconductivity_2001, PhysRevLett.95.087003, PhysRevB.87.241408, PhysRevB.90.245105}. The resulting interplay of interactions, which are effective at different length and time scales (see Fig. \ref{fig:introduction}), makes the phase diagrams of two-dimensional materials and their response to external stimuli very rich. 
	
	The layered metallic transition metal dichalcogenides (TMDC)~\cite{wang_electronics_2012,manzeli_2d_2017}, MX$_2$, where M denotes one of the transition metals V, Nb, or Ta and X stands for one of the chalcogens S or Se, presents a demonstrative case in this respect where the monolayer limit is becoming experimentally accessible~\cite{wang_electronics_2012,xi_strongly_2015,manzeli_2d_2017}. Within this material class a competition of charge- and spin-ordered, Mott insulating as well as superconducting states can be found. Here, the V-based compounds show tendencies towards magnetic~\cite{zhuang_stability_2016,isaacs_electronic_2016} as well as charge order~\cite{PhysRevB.82.075130, C5CP01326G, PhysRevB.89.235125} in their monolayer and bulk phases, respectively, which might partially coexist in the few-layer limit~\cite{ANIE:ANIE201304337, isaacs_electronic_2016}. In contrast, the sub-class of Ta-based compounds~\cite{pillo_interplay_2000, cho_interplay_2015, cho_nanoscale_2016, ma_metallic_2016, sipos_mott_2008, liu_electron-phonon_2009, yu_gate-tunable_2015} as well as NbSe$_2$~\cite{PhysRevB.92.140303,calandra_effect_2009,leroux_anharmonic_2012,flicker_charge_2015,flicker_charge_2015-1,flicker_charge_2016,chatterjee_emergence_2015,ugeda_characterization_2016, xi_strongly_2015,nakata_monolayer_2016} show a competition between charge-density waves, superconducting as well as Mott insulating states. NbS$_2$ appears to be a border case. It is superconducting in the bulk~\cite{guillamon_superconducting_2008,tissen_pressure_2013,Heil17} but does not display any charge-density wave formation there. In the case of few-layer NbS$_2$ first experimental studies reported recently metallic transport properties down to three layers~\cite{zhao_two-dimensional_2016}, while mean-field calculations reveal a tendency to form magnetic states~\cite{xu_tensile_2014,guller_spin_2016}. NbS$_2$ is thus likely on the verge between different instabilities. Whether or how these instabilities are triggered by the interplay of the involved interactions is barely understood, up to now. While a lot of focus was put on the investigation of electron-phonon coupling effects in the whole class of metallic TMDCs, the effects of the subtle interplay of the local and non-local Coulomb interaction terms have been mostly neglected, so far. It is thus necessary to draw our attention also to these short and long-range electron-electron interactions. Unfortunately, there is no theory that can handle, even qualitatively, the competition of these strong interactions beyond the perturbative regime, yet. 
	To overcome this problem, we combine here the Dual Boson formalism~\cite{Rubtsov12,vanLoon14,Hafermann14-2} with first-principles approaches and construct a state-of-the-art material-realistic theory of monolayer NbS$_2$ which properly treats electronic correlations as resulting from competing short- and long-range Coulomb interactions. Thereby, we also account for the electron-phonon interactions to gain a universal understanding of all interaction effects.

	Our calculations reveal a simultaneous enhancement of the charge and spin susceptibilities due to the various interactions in monolayers of NbS$_2$ and a sharp transition from tendencies of preferential spin ordering to charge ordering. Despite these strong interaction effects, the electronic spectral function as measured, e.g., in angularly resolved photoemission (ARPES) experiments largely resembles the non-interacting dispersion in accordance with the available experimental data. We trace this back to a compensation of the different interaction terms which are partially effective on the single-particle but not on the two-particle level.
	From an experimental perspective, this means that finding a match between ARPES results and DFT bands is not sufficient to rule out strong correlation/interaction effects, since the competition of the interactions masks correlation effects on the single-particle level while they are still visible on the two-particle level, i.e., in the magnetic and charge susceptibility.

	\section{Results}
	
	\textbf{Competing interactions in NbS$_2$.} The non-correlated band structure of NbS$_2$ monolayers exhibits a half-filled metallic band surrounded by completely filled valence bands ${1}\,$eV below and completely empty conduction bands $3\,$eV above the Fermi level, as shown in Fig.~\ref{fig:introduction} (a). This motivates a description of the competing interaction effects in terms of an extended Hubbard-Holstein model~\cite{berger_two-dimensional_1995} for the separated metallic band only
	\begin{align}
		H^\text{sb} =
		 	& -\sum_{i,j} \sum_\sigma t_{ij} c_{i\sigma}^\dagger c_{j\sigma} \nonumber\\
		 	& + U \sum_{i} n_{i\uparrow} n_{i\downarrow}
		+ \frac{1}{2}\sum_{\substack{ i \neq j \\ \sigma \sigma'}} V_{ij} n_{i\sigma} n_{j\sigma'} \nonumber\\ 
		 	& + \omega_{\text{ph}} \sum_{i} b_i^\dagger b_i + g\sum_{i \sigma} n_{i\sigma} (b_i+b_i^\dagger),
		 	\label{eq:ModelHamiltonian}
	\end{align}		
	where $c^\dagger_{i\sigma}$ and $c_{i\sigma}$ are the creation and annihilation operators of the electrons with spin $\sigma$ on lattice site $i$, $b^\dagger$ and $b$ are the creation and annihilation operators of a local phonon mode, and $n_i=c^\dagger_{i\sigma}c_{i\sigma}$ is the electron occupation number operator. This  model includes on the single-particle level the electron hopping $t_{ij}$ and a local phonon mode with energy $\omega_\text{ph}$. We include an on-site Coulomb repulsion $U$ and long-range Coulomb interactions $V_{ij}$, as well as an electron-phonon coupling $g$ which couples the local charge density to the given phonon mode. The latter can actually be integrated out which results in a purely electronic model with an effective dynamic local interaction 
	\begin{align}
		U \rightarrow U_\text{eff}(\omega) = U - \frac{2g^2 \omega_{\text{ph}}}{\omega_{\text{ph}}^2 - \omega^2}
		\label{eq:UEff}
	\end{align}
	that is lowered and thereby effectively screened by the phonons~\cite{berger_two-dimensional_1995, PhysRevLett.94.026401, PhysRevLett.99.146404}. This treatment of the phonons as simple single-frequency modes that are coupled locally to the electrons is an assumption necessary to keep the problem tractable. Otherwise, the non-local interaction $V_{ij}$ or $V_q$ in momentum space would also become frequency dependent. As we argue in the Methods section, there is a basis for this assumption, however, the simplification might change the exact position at which instabilities occur in the Brillouin zone. Furthermore, we would like to emphasize that, in our treatment, electronically generated phonon anharmonicities are automatically included whereas bare phonon anharmonicities are not.
		
	To realistically describe NbS$_2$ monolayers, we derive the parameters entering Eq.~(\ref{eq:ModelHamiltonian}) from first principles. Therefore, we generate a most accurate tight-binding model describing the metallic and the lowest two conduction bands in a first step, and use it afterwards to perform calculations within the constrained Random Phase Approximation (cRPA)~\cite{PhysRevB.70.195104} to obtain the partially screened Coulomb interaction matrix elements within the same basis. The phonon frequency and the electron-phonon coupling are estimated based on density functional perturbation theory calculations. The resulting three-band model is subsequently simplified in order to get the final single-band model describing the metallic band only as explained in the method section.

	Our \abinitio\ simulations yield an effective local Coulomb interaction $U\approx1.8\,$eV, a nearest-neighbor interaction $V\approx1\,$ eV as well as further long range interaction terms. The typical bare phonon frequency $\omega_{ph}$ and the electron-phonon coupling $g$ for this material are estimated to be $20\,$meV and $70\,$meV, respectively (see methods for further details). Notably, both, the on-site Coulomb repulsion and the effective electron-electron attraction $\lambda=2g^2/\omega_{ph}=0.5\,$eV, are on the order of the electronic band width $\approx 1.2\,$eV, as sketched in Fig. \ref{fig:introduction}. 

	\begin{figure}[hbt]
		\centering
		\includegraphics[scale=0.67]{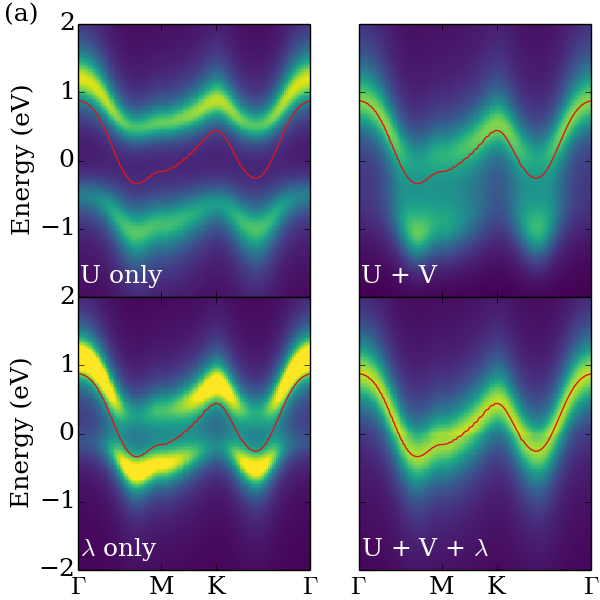}
		\includegraphics[scale=0.67]{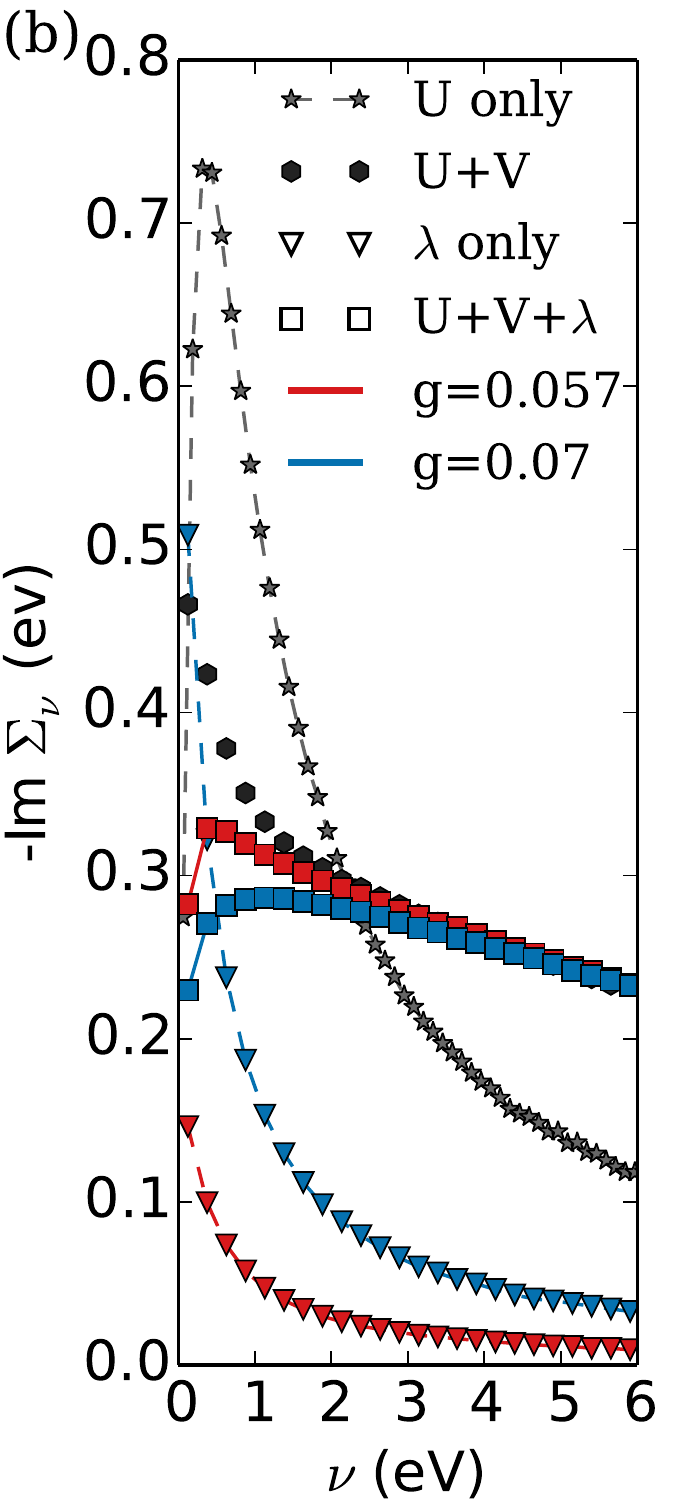}
		\caption{Spectral fingerprints of competing interactions.  
			(a) Momentum and energy resolved spectral functions as obtained with the Dual Boson method involving the local Coulomb interaction only (top left), the full Coulomb interaction (top right), the electron-phonon interaction $\lambda$ with $g=0.07$\,eV (bottom left) and all interaction terms (bottom right). The red line indicates the bare metallic band as obtained from DFT and shown in Fig.~\ref{fig:introduction} (a). The top panel is calculated at $232\,$K, the other panels are calculated at $464\,$K. (b) The corresponding local self-energies as functions of Matsubara frequencies. Results for two different electron-phonon coupling strengths $g_1=0.057$\,eV and $g_2=0.07$\,eV are shown in red and blue respectively. 
		}
		\label{fig:specfunc}
	\end{figure} 
 	
	\textbf{Spectral Fingerprints of the Interactions.} Each interaction term on its own can thus trigger strong electronic correlations, which becomes evident from the electronic spectral functions shown in Fig. \ref{fig:specfunc} (a). 
	These have been calculated using the Dual Boson (DB) method taking into account each interaction term on its own and their combined effects.
	For local Coulomb interactions only (top left), the half filled conduction band clearly splits into two Hubbard bands above and below the Fermi level. There is no spectral weight at the Fermi level and the system is insulating. Including the non-local Coulomb interaction terms (top right) markedly changes the spectral function. The lower Hubbard band still retains noticeable spectral weight. However, the upper Hubbard band overlaps now with a broad distribution of spectral weight reaching the Fermi level. That is, the non-local Coulomb interaction drives the system into the state of a correlated metal, similar to what has been shown for graphene \cite{PhysRevLett.111.036601}.
	With only electron-phonon interaction (bottom left), the spectrum is also reminiscent of a correlated metal, with again strong spectral weight transfer away from the Fermi level towards polaronic bands at higher energies. 
	Finally, simultaneous inclusion of all interactions yields the spectral function shown in the bottom right of Fig. \ref{fig:specfunc} (a). We find a single band with a dispersion very similar to the DFT result of Fig.~\ref{fig:introduction} (shown as a red line). Seemingly, the different interaction terms largely compensate each other despite the fact that they are effective at very different length and time scales. The major interaction effect visible in, e.g., ARPES experiments is that the band widens significantly compared to the thermal broadening inherent to any finite temperature measurement. ARPES experiments in other transition metal dichalcogenides monolayers (TaS$_2$~\cite{Sanders16} and NbSe$_2$~\cite{ugeda_characterization_2016}) are consistent with this picture: the dispersion follows roughly the DFT band structure with some broadening. A more detailed comparison with our results is, however, not possible, since in the experiments different materials have been used, lower temperatures were applied, and substrates were present. 
	
	Our material-specific results can be compared to theoretical findings in model systems. 
	In the Hubbard-Holstein model ($U$ and $\lambda$ in our language) on the triangular lattice, Mott and polaronic insulating states have been found~\cite{Jeon04,Yoshioka10}, consistent with our results here. On the other hand, the combination $U+V+\lambda$ presented here has so far not been studied since there were previously no methods that can deal with these competing interactions in the strongly correlated regime, where vertex corrections beyond $GW$ are important.

	Therefore, we need to scrutinize this behavior in more detail and examine the local self-energy, which induces all correlation effects. In Fig.~\ref{fig:specfunc} (b), we show these self-energies corresponding to the spectra in Fig.~\ref{fig:specfunc} (a). If we take only local Coulomb interactions into account (stars), we find a strongly enhanced self-energy for small frequencies. By including also long-range Coulomb interactions (circles), the self-energy is reduced around small frequencies. This trend is continued by including electron-phonon interactions (squares), which demonstrates how the long-range Coulomb and the electron-phonon interactions compensate the effects of the local Coulomb interaction. The full self-energy including all interaction terms is thus strongly reduced around small frequencies, but has still sizeable contributions at all energies considered here, which results in the broadened spectral function without significant reshaping. It is interesting to note that when taking only the electron-phonon interactions into account (triangles), the self-energy, and thus the degree of correlation increases by increasing the electron-phonon coupling $g$. However, in the presence of Coulomb interaction an enhanced electron-phonon coupling necessarily leads to a decrease of the self-energy and hence to a decreased degree of correlation.
	Thus, the effect of electron-phonon coupling is the exact opposite depending on whether or not the Coulomb interaction is present  in the model. 
	It is therefore absolutely crucial to take all interactions simultaneously into account. 

	\begin{figure}
	\centering
	\includegraphics[width=1.0\columnwidth]{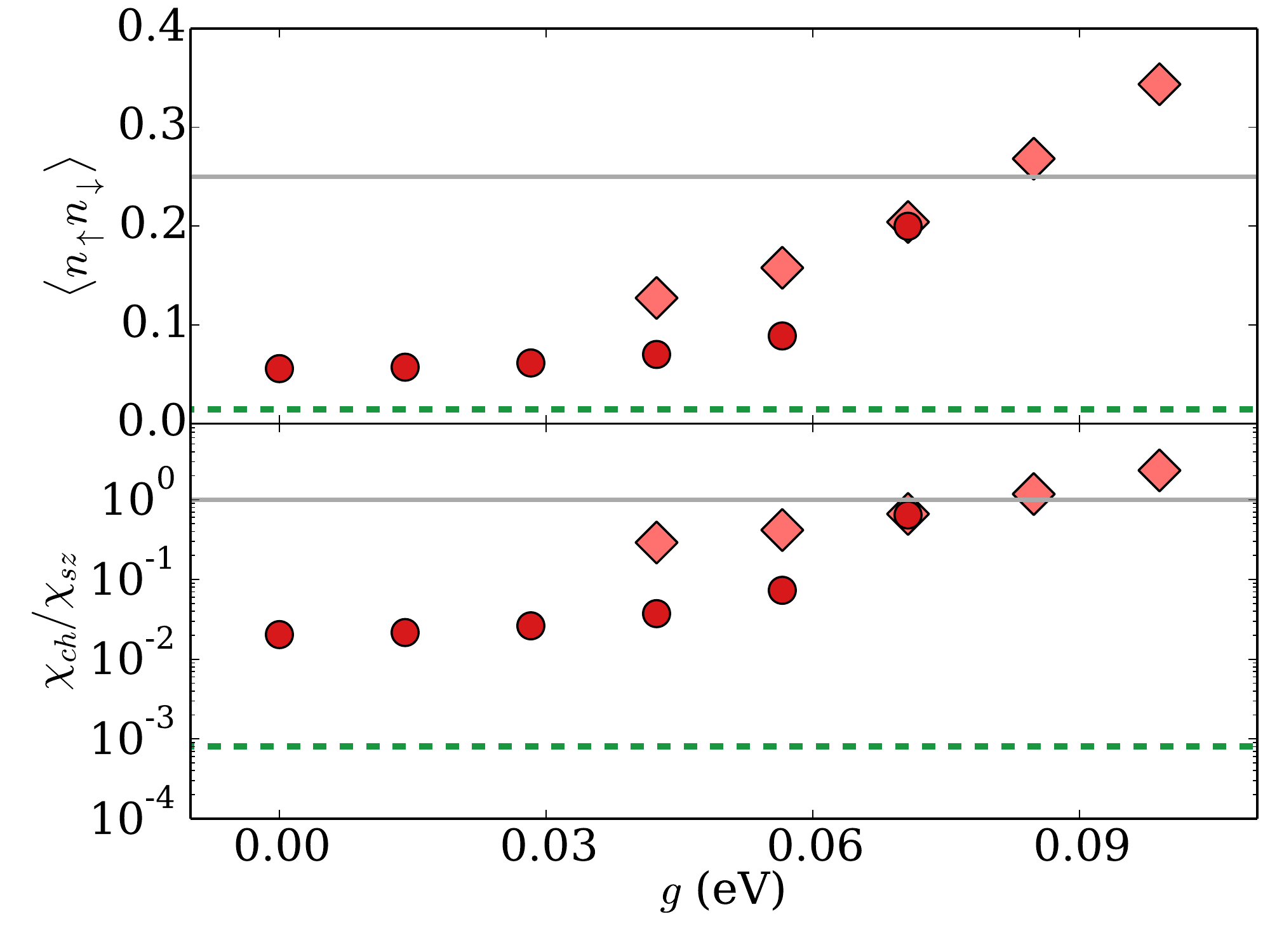}
	\caption{Double occupancy ($\braket{n_\uparrow n_\downarrow}$, upper panel) and ratio of local 	charge and spin susceptibility ($\chi_{ch}/\chi_{sz}$, lower panel) as function of the electron-phonon coupling strength $g$. Circles are obtained from the auxiliary impurity model of the Dual Boson simulations at $T=464\,$K, diamonds at $T=2321\,$K. Dashed green lines represent data for purely local interaction $U$; solid grey lines corresponds to the non-interacting limit.
		}
		\label{fig:AuxSystem}
	\end{figure}

	\begin{figure*}
		\begin{minipage}[t]{0.32\textwidth}
			\includegraphics[width=\textwidth]{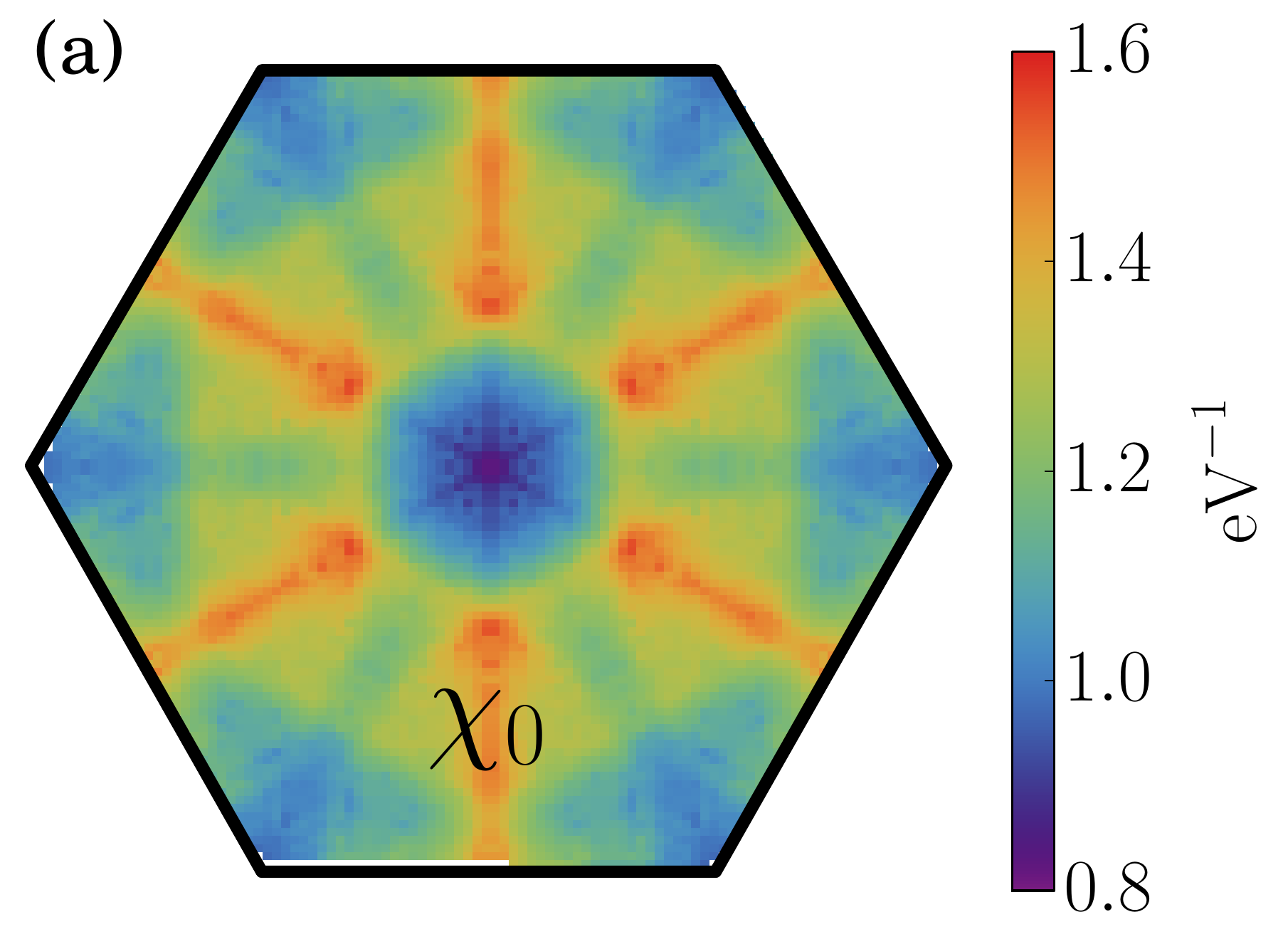} %\\
		\end{minipage}
		\begin{minipage}[t]{0.32\textwidth}
			\includegraphics[width=\textwidth]{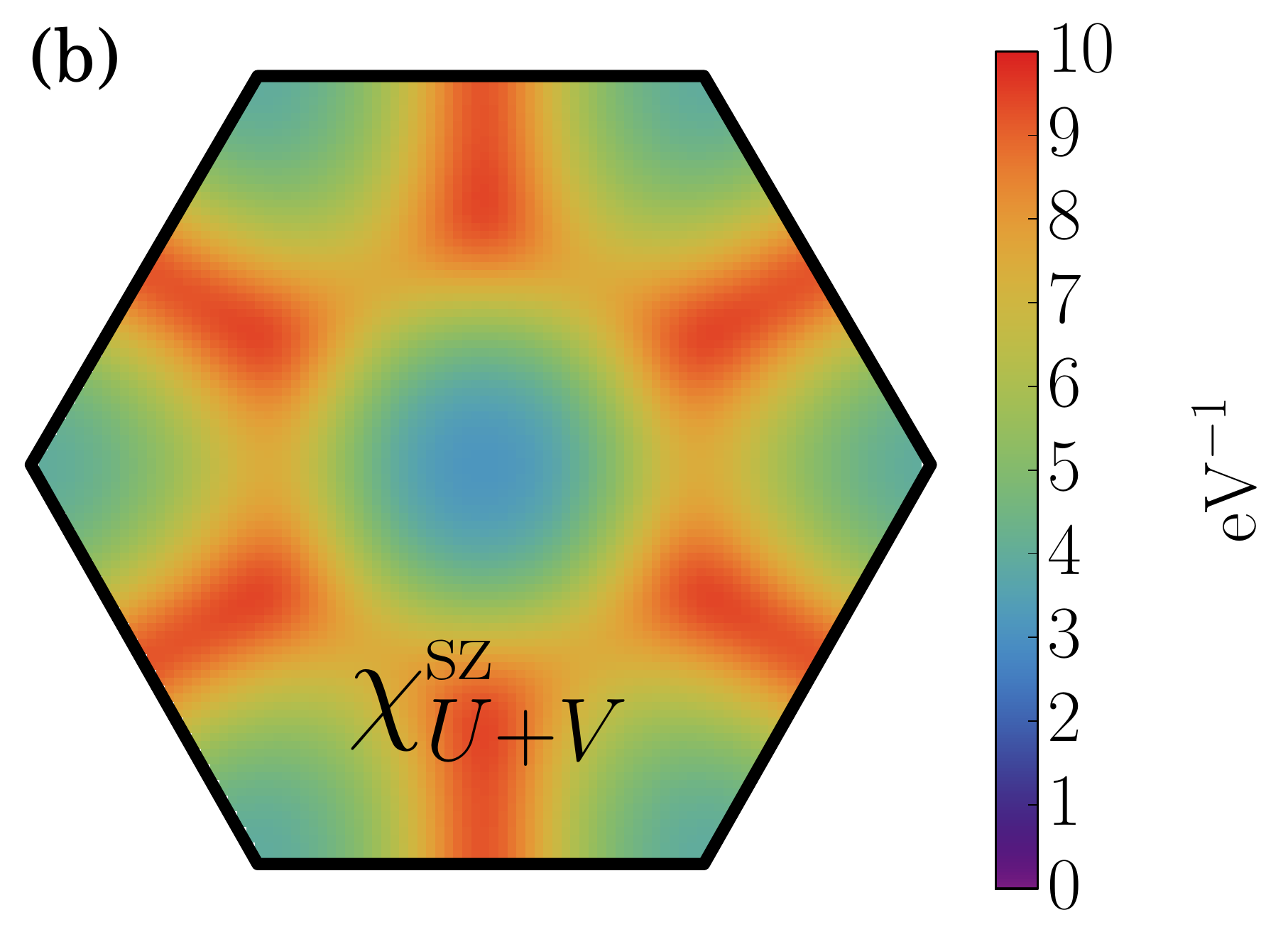} %\\
			\includegraphics[width=\textwidth]{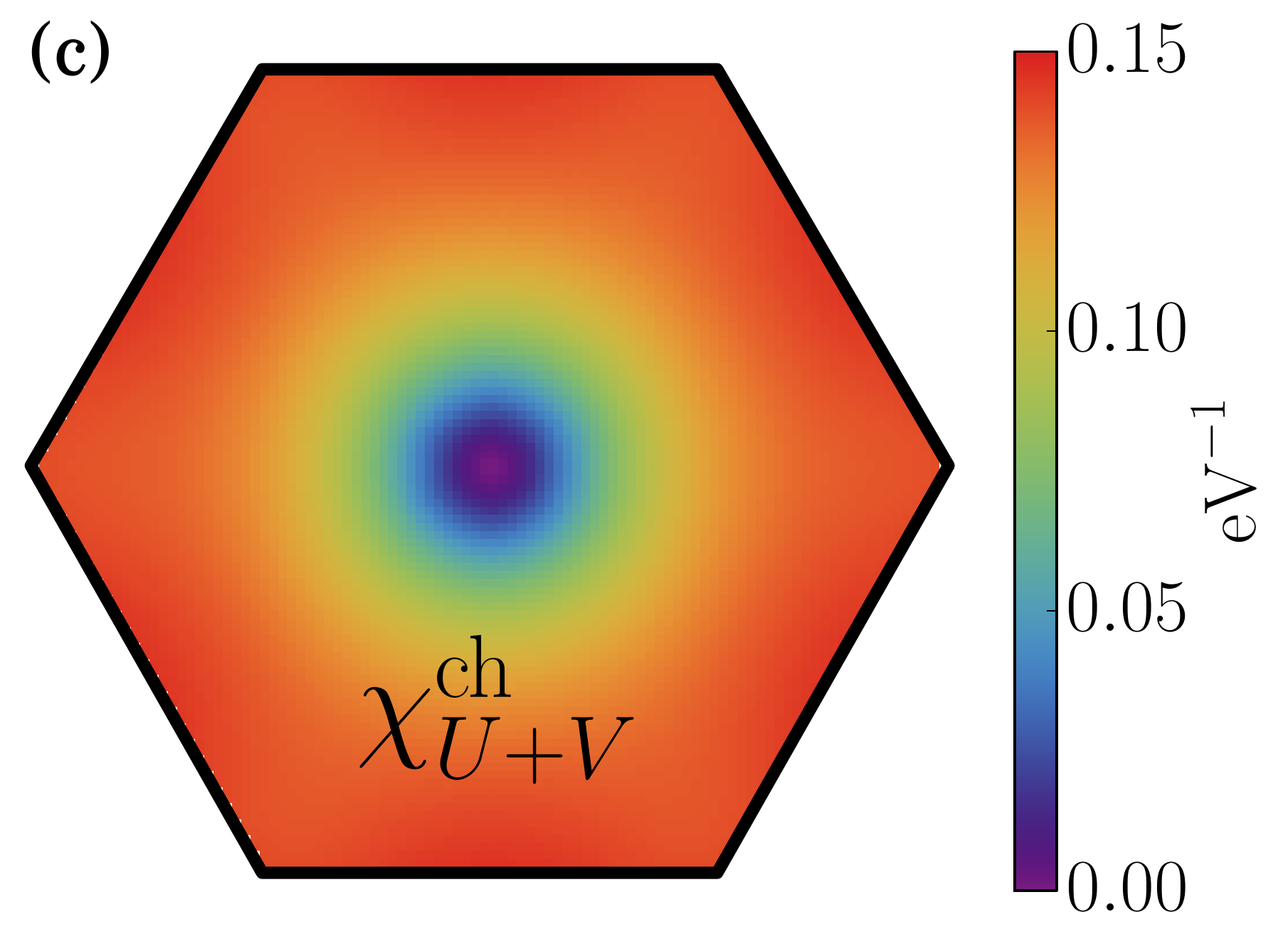} %\\
		\end{minipage}
		\begin{minipage}[t]{0.32\textwidth}
			\includegraphics[width=\textwidth]{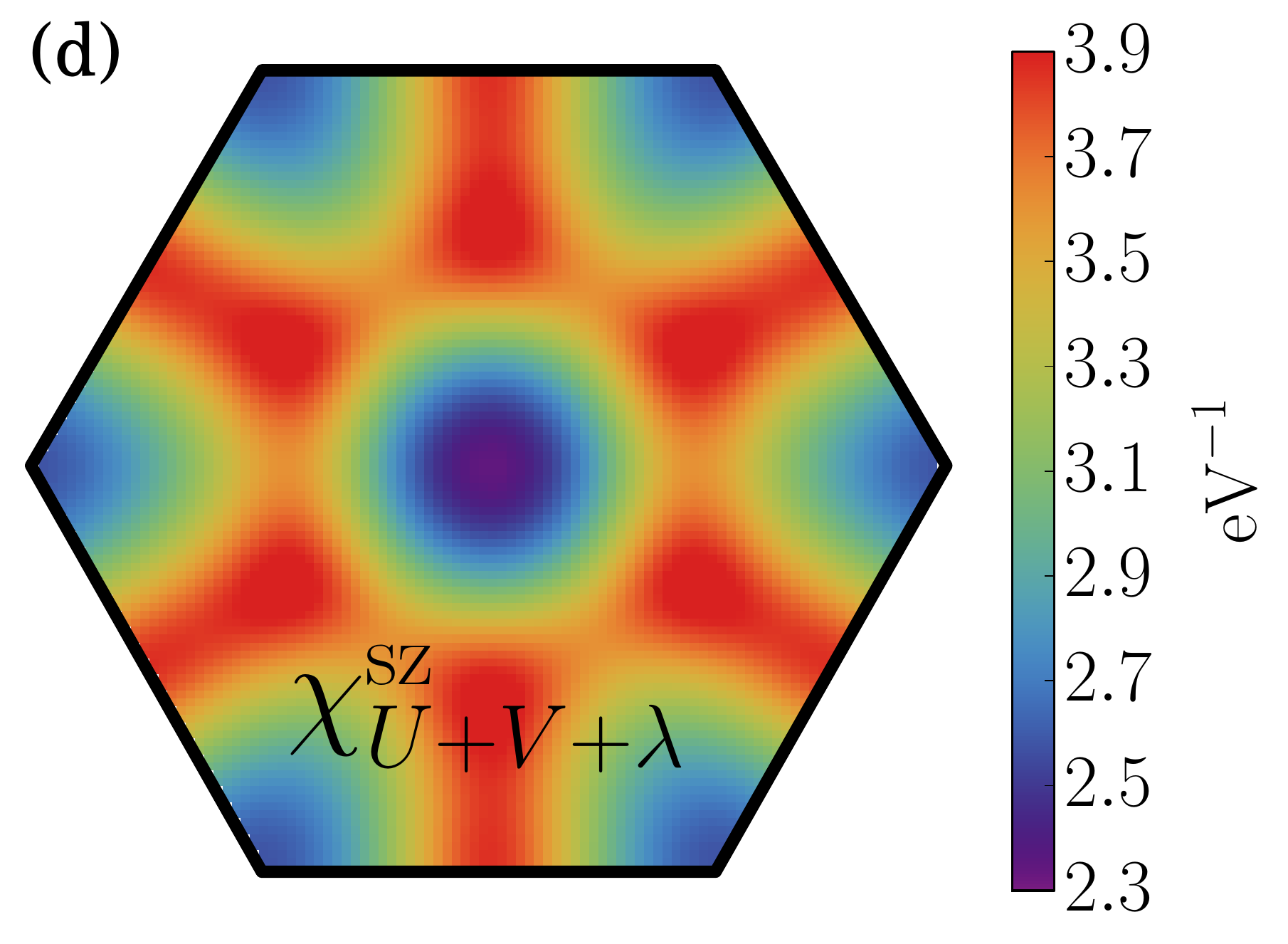} %\\
			\includegraphics[width=\textwidth]{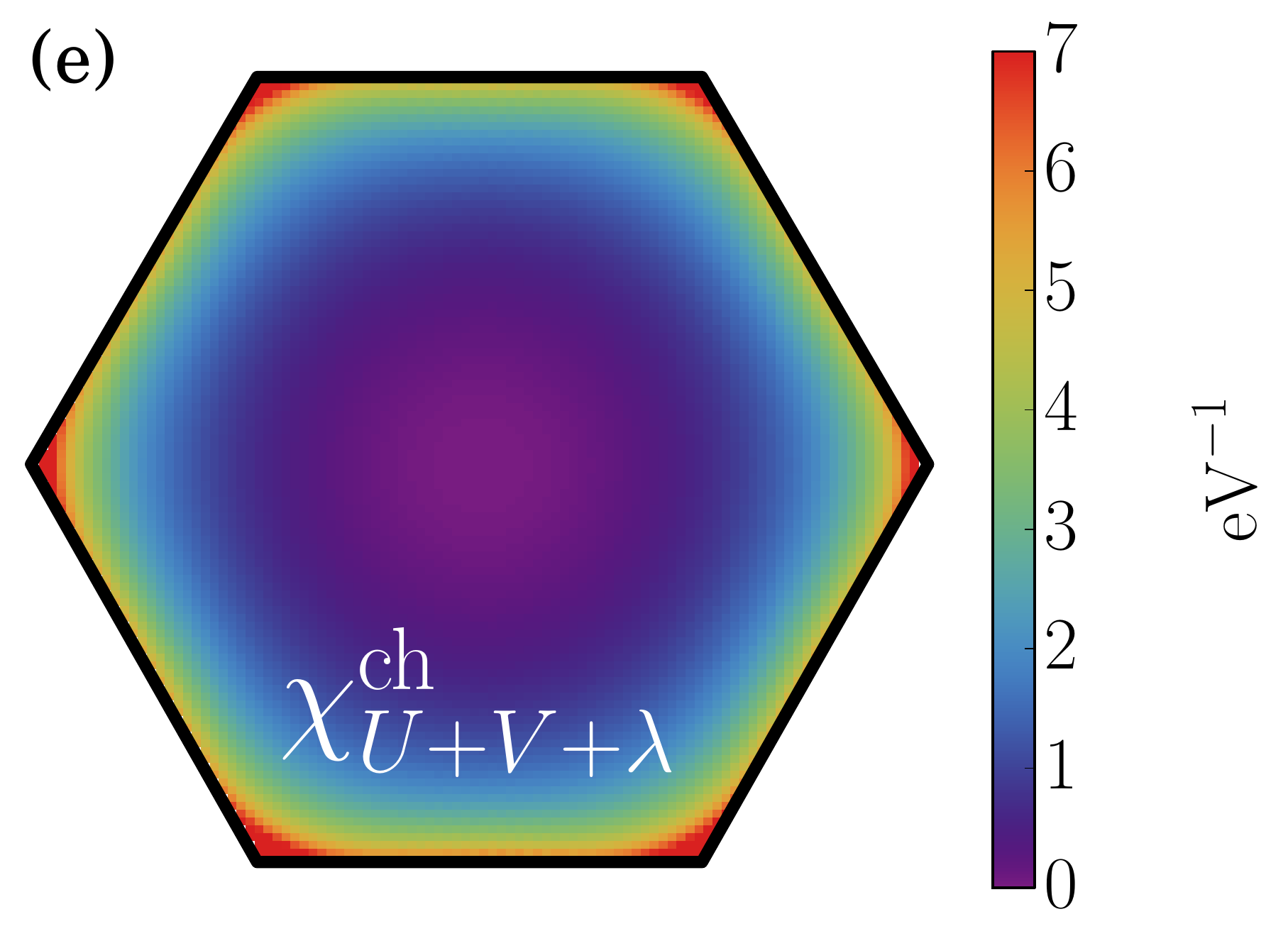} %\\
		\end{minipage}
		\caption{Static susceptibilities as a function of momentum in the Brillouin Zone. (a) Non-interacting susceptibility. (b) and (c) Interacting spin and charge susceptibility for $g=0$. (d) and (e) Interacting spin and charge susceptibility for $g=70$\,meV, $T=464\,$K.\label{fig:DBSuceptiilities}}
	\end{figure*}	

	\begin{figure}
		\centering
		\includegraphics[width=1.0\columnwidth]{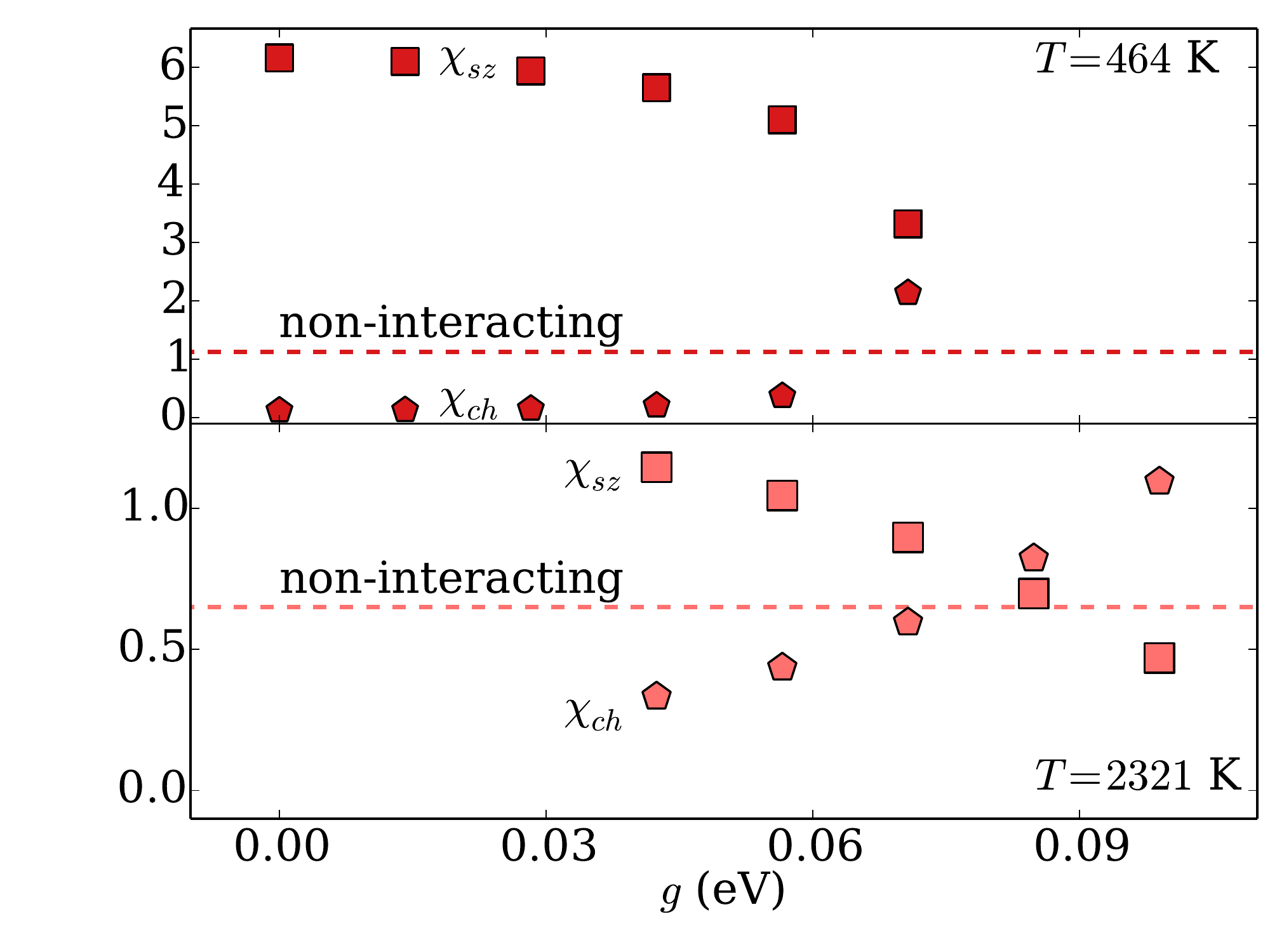}
		\caption{Local charge (pentagons) and spin (squares) susceptibility as function of the electron-phonon coupling strength $g$, c.f. \ref{fig:AuxSystem} for the ratio of susceptibilities. The two panels represent $T=464\,$K and $T=2321\,$K. The dashed lines represent the susceptibility of the non-interacting system.
		}
		\label{fig:localsusc}
	\end{figure}		
	
	\textbf{Competition of charge and spin fluctuations in NbS$_2$ monolayers.} The electronic correlations as resulting from the interplay of the electron-electron and electron-phonon interaction also manifest in the local two-particle correlation functions of the system, which are shown as a function of the electron-phonon coupling $g$ in Fig.~\ref{fig:AuxSystem}. These local observables are calculated directly from the DB auxiliary single-site system. The ratio of the static local charge and spin susceptibilities (lower panel; note the logarithmic scale) and the instantaneous double occupancy (upper panel) vary strongly as a function of $g$. Without electron-phonon coupling ($g=0$) the system shows typical signs of strong Mott-Hubbard correlation effects: The spin susceptibility is orders of magnitude larger than the charge susceptibility and the probability of finding two electrons at the same site is greatly reduced in comparison to the value of $\braket{n_\uparrow n_\downarrow} = 0.25$ found in non-interacting half-filled systems. Turning on the electron-phonon interaction screens the local Coulomb interaction according to Eq.~(\ref{eq:UEff}), and makes the system less correlated. At sufficiently large $g \approx 70..80\,$meV, the susceptibility ratio and double occupancy even exceed their non-interacting values of $1$ and $0.25$ (grey lines), respectively.
	The numerical simulations get unstable close to a transition to the charge-density wave (CDW) phase. This is why we could perform simulations at $464\,$K only up to $g=70\,$meV (red circles). At a higher temperature of $T=2321\,$K (orange pluses), larger values of $\lambda$ can be reached. 
	For $g=40-70\,$meV, the two data sets agree reasonably well, at higher temperatures the double occupancy is less suppressed and the spin susceptibility is substantially smaller, see also the methods section.
	Note that $464\,$K ($0.04\,$eV) is well below the energy scales defined by the band width and the interactions.

	Freestanding NbS$_2$ monolayer, with $g\approx 70$\,meV as estimated in the methods, thus turn out to be on the verge to form a charge-density wave ground state.
	The local properties presented in Fig.~\ref{fig:AuxSystem} also show what happens when both, the electron-phonon interaction and the non-local parts of the Coulomb interaction, are ignored. In that case (dashed green lines), the susceptibility ratio goes down another order of magnitude, and the double occupancy decreases to almost zero. These are all characteristics of a Mott insulating phase. The local Hubbard interaction $U$ is thus in principle strong enough to create an interaction-driven insulator, with a large spin susceptibility, local magnetic moments (small $\braket{n_\uparrow n_\downarrow}$) and strongly suppressed charge fluctuations, as we have already anticipated in the discussion of the spectral functions. Only through screening by the non-local Coulomb contributions, and by the electron-phonon coupling, can the system exhibit the large charge fluctuations (local ``charge moments'', large $\braket{n_\uparrow n_\downarrow}$) that are necessary for a charge-density wave. This shows that both, the Hubbard interaction $U$ and the interactions that screen it, are non-perturbatively large in the freestanding monolayer, which casts doubt on approaches that do not explicitly include all interaction terms. Most importantly, the transition from the regime which is dominated by spin-fluctuations to the charge-fluctuation dominated regime is very abrupt as the steep rise of the susceptibility ratio demonstrates. The strong fluctuations in different channels, around $g\approx70\,$meV, signal the close proximity of competing charge and spin order and is indeed ubiquitous in correlated electron systems~\cite{Tranquada95,Hansmann13}.

	Next, we turn to the static momentum-resolved susceptibilities. The non-interacting susceptibility of the single-band model, $\chi_0$, shown in Fig.~\ref{fig:DBSuceptiilities} (a), agrees with previously published data for NbS$_2$ monolayers~\cite{guller_spin_2016}. In a non-interacting system the charge and spin susceptibility would be the same and coincide with $\chi_0$. This is clearly not the case for the charge and magnetic susceptibilities resulting from our DB calculations shown in  Fig.~\ref{fig:DBSuceptiilities} (b)-(e). 

	Without electron-phonon coupling ($g=0$) the spin susceptibility is enhanced indicating the presence of strong spin fluctuations. The charge susceptibility, on the other hand, is suppressed in the entire Brillouin zone due to the Coulomb interaction, which is in line with the expectations for a correlated metal. Turning on the electron-phonon interaction ($g=70$\,meV) reduces the spin susceptibility, which is however still comparable to $\chi_0$. At the same time, the charge susceptibility is strongly enhanced and is almost divergent at large momenta. At this point it is, however, important to note that the exact position of the ordering vector might change when the ordered phase is actually entered (here, we investigate just its onset based on the susceptibility in the normal phase) and when a more realistic phonon model is used. Nevertheless, these two observations show again one of our main findings that the interactions partially compete and screen each other, leading to a spin susceptibility that is only moderately enhanced. Most importantly, this competition does not lead to a complete cancellation, as is visible in the strong enhancement of the charge susceptibility. Due to the interplay of these interactions a strong spin and charge response can thus coexist in this system. 

	\section{Discussion}
	
	Using a combination of the Dual Boson approach and \abinitio\ calculations, we investigated the interplay between the Coulomb and electron-phonon interactions in NbS$_2$ monolayers and the resulting degree of electronic correlations. We found that both, the Coulomb and the phonon-mediated electron-electron interaction energies, are on the same order as the electronic band width allowing both of them to trigger strong electronic correlations. Both types of interactions on their own would drive NbS$_2$ to the verge of an insulating state, as our analysis of electronic self-energies shows. Remarkably, our simulations with Coulomb and electron-phonon interactions present yield a spectral function which closely resembles the non-interacting band structure. Yet, in this situation electron correlations have not fully ceased but manifest themselves in a sizeable broadening of the spectral function. In this sense NbS$_2$ is very similar to so-called Hund's metals~\cite{Werner08,Haule09,Yin11,Medici11,Georges13}, where the exchange coupling drives the electronic system of materials like Fe-based superconductors away from the Mott Hubbard insulating limit into a correlated metallic phase.
	
	For NbS$_2$, the observed spectral broadening argues against simple nesting scenarios based on the bare bands for the CDW instabilities. Our findings rather show that the interplay between all interactions is responsible for driving the system in close proximity of a charge-ordered state.
	The interaction-induced correlations result in strongly modified spin and charge susceptibilities compared to the non-interacting one. Specifically, we found that the competition between the long range Coulomb and the electron-phonon interactions is responsible for NbS$_2$ monolayers being on the edge between dominating spin- and charge fluctuations. The transition from a preferential spin order to charge order is thereby abruptly driven by the electron-phonon interaction. 
	
	The resulting ground state is thus heavily dependent on the detailed balance between the internal interactions. To study and test this behavior experimentally, there are several points to be aware of. First, from matching ARPES and DFT data one can not deduce that the mean-field calculation captures the main physics. Many-body effects lead to quasiparticle broadening as well as enhanced magnetic and charge susceptibilities, which can be measured directly in resonant x-ray scattering~\cite{Ament11} or electron energy loss spectroscopy~\cite{Vig17}. Importantly, the predicted close vicinity of charge order and local magnetic moment formation can be experimentally tested. The electronic system of NbS$_2$ can be manipulated via environmental screening or strain applied to the monolayer. Increasing the former will mostly reduce the long-range Coulomb interaction $V$ while the electron-phonon interaction remains largely untouched. Thereby, the effective screening of the local $U$ due to the non-local $V$ is reduced and the spin susceptibility should be enhanced. By applying strain the electron-phonon interaction can be varied without drastic changes to the Coulomb interaction. By increasing the electron-phonon interaction the system would be pushed into a charge ordered state. 
	
	Our calculations show that correlation effects are particularly prominent in the simultaneously enhanced spin and charge susceptibilities of NbS$_2$. 
	Hence, we expect a strong response of the material to local perturbations, which can be experimentally realized through charged as well as magnetic adsorbates. Scanning tunneling microscopy experiments involving adsorbates or defects on NbS$_2$, similar to what has been done for NbSe$_2$~\cite{Menard15,Kezilebieke18}, allow the susceptibilities predicted in Fig.~\ref{fig:DBSuceptiilities} to be probed in real space but could possibly also exploit NbS$_2$ as a platform for quantum engineering, where one switches locally between spin and charge order. The combination of the lattice structure of NbS$_2$, the sizeable spin-orbit coupling, and the enhancement of spin and charge susceptibilities clearly away from the Brillouin zone center suggests that the competing ordering tendencies is likely subject to frustration effects. For these reasons, monolayer NbS$_2$ deserves future exploration not only in the light of fundamental interest but possibly also in relation to concepts of miniaturized neuromorphic computing~\cite{Engel2001}.
	
	Finally, our findings allow to speculate about possible superconducting properties. In this context, the enhanced spin and charge susceptibilities point towards interesting unconventional paring mechanisms. At the same time there are the before mentioned striking similarities between the self-energy in NbS$_2$ and the one found in Hund's metals, yielding a similar scenario as in Fe-based superconductors. Additionally, it needs to be pointed out that the appearance of a CDW phase is usually detrimental to superconductivity so that it likely needs to be suppressed to enhance T$_c$. Given the complicated competition between magnetic and charge instabilities in NbS$_2$, an analysis including all of these aspects needs to be carried out to gain reliable insights into possible superconducting properties.
	
	\section{Methods}
	
	\textbf{Parametrization of the Extended Hubbard-Holstein Model.} All model parameters are derived from first principles based on DFT and constrained random phase approximation (cRPA) calculations. To do so, we start with a DFT calculation in Fleur \cite{FLEUR} for NbS$_2$ using a lattice constant of $a = 3.37$\,\AA, a $k$ mesh of $18 \times 18 \times 1$, a vacuum height of $32$\,\AA, a relaxed sulfur-sulfur distance of $\Delta = 3.13$\,\AA, and using FLAPW l-expansion cutoffs of 10 (Nb) and 8 (S) and muffin tin radii of $2.58\,$a$_0$ (Nb) and $2.01\,$a$_0$ (S) to calculate the band structure shown in Fig. \ref{fig:introduction} (a). 
	Since the spin-orbit coupling leads to severe spin splittings at the $K$ point only, but not around the Fermi level in NbS$_2$ we neglect it in the following. Afterwards we construct a three-band tight-binding model by projecting the original DFT wave functions onto the three dominant niobium orbitals ($d_{z^2}$, $d_{xy}$, $d_{x^2-y^2}$) using the Wannier90 code\cite{mostofi_updated_2014}, whereby we ensure that the bands are properly disentangled. To preserve the orbital character of the Wannier functions, we do not perform maximal localization. The resulting three-band tight-binding model perfectly interpolates the original DFT band structure and can be used to evaluate the electronic dispersion at arbitrary $k$ points.
	
	The long-range Coulomb interaction is parametrized in a material-realistic manner using the cRPA method \cite{PhysRevB.70.195104}. Therefore, we start with the fully screened dynamic Coulomb interaction $W(q, \omega)$ in reciprocal space which is defined by
 	\begin{align}
		W(q,\omega) = \frac{v(q)}{1 - \Pi(q, \omega) v(q)},
	\end{align}
	where $v(q) \propto 1/q$ is the \textit{bare} interaction in two dimensions and $\Pi(q,\omega)$ is the polarization function rendering all screening processes. According to the cRPA we can reformulate the latter $\Pi(q, \omega) \approx \Pi_\text{mb}(q, \omega) + \Pi_\text{rest}(q)$ by splitting it into a \textit{dynamic} part arising from the half-filled metallic band (mb) and a \textit{static} part resulting from the rest of the band structure. This is appropriate since we are interested in the low-frequency properties of $\Pi(q, \omega)$ and $W(q, \omega)$ only, which are completely rendered by the metallic band and thus by $\Pi_\text{mb}(q, \omega)$. Using this formulation of the full polarization we can rewrite the fully screened interaction as follows
	\begin{align}
		W(q, \omega) = \frac{U(q)}{1 - \Pi_\text{mb}(q, \omega) U(q)}
		\label{eq:WcRPA}
	\end{align}
	with $U(q)$ being the \textit{partially screened} Coulomb interaction defined by
	\begin{align}
		U(q) 
			= \frac{v(q)}{1 - \Pi_\text{rest}(q) v(q)}
			= \frac{v(q)}{\varepsilon_{rest}(q)}.
		\label{eq:UcRPA}
	\end{align}
	As described in the supplemental methods, $U(q)$ needs be evaluated within the same orbital basis as used for the tight-binding dispersions, using $3 \times 3$ matrices to represent the bare interaction $\mathbf{v}(q)$ and the dielectric function $\mathbf{\varepsilon}(q)$. Importantly, we can fit analytic expressions to all of the involved matrix elements $U_{\alpha \beta}(q)$ allowing us to evaluate $\mathbf{U}(q)$ at arbitrary $q$ vectors.
	
	In order to derive a single-band model we neglect the orbital dependencies in a next step. In this case the dynamic polarization \textit{matrix} of the metallic band may be approximated via  
	\begin{align}
		\mathbf{\Pi}_\text{mb}(q, \omega) = 
		\frac{1}{9} \
		\Pi_\mathrm{sb}(q, \omega)
		\begin{pmatrix}
		1 & 1 & 1 \\
		1 & 1 & 1 \\
		1 & 1 & 1
		\end{pmatrix},
	\end{align}
	where $\Pi_{sb}(q, \omega)$ is the single-band polarization which is going to be evaluated in the Dual Boson calculations. The factor $\frac{1}{9}$ approximates the overlap matrix elements which are in general orbital and momentum dependent. This is appropriate for small $q$ and as long as all orbital weights are more or less the same. We found that this assumption is indeed valid in the half-filled situation discussed here. Using this polarization corresponds to a single-band/orbital partially screened Coulomb interaction defined by
	\begin{align}
		U_\mathrm{sb}(q) =
		\frac{1}{9}
		\sum_{\alpha \beta} 
		U_{\alpha \beta}(q).
		\label{eq:SingleBandCoulombInteraction}
	\end{align}
	Thus, under the assumption of vanishing orbital dependencies we can define the partially screened Coulomb interaction of the single-band model as the arithmetic average of all matrix elements of the partially screened interaction matrix $\mathbf{U}(q)$ in the orbital basis. This $U_\mathrm{sb}(q)$ now represents the Fourier transform of the real-space Coulomb interactions $U$ and $V$ as used in Eq.~(\ref*{eq:ModelHamiltonian}) and thus serves as the second important ingredient to our extended Hubbard-Holstein model.
	
	Finally, we incorporate the phonon frequency and the electron-phonon coupling into our model to describe all important interactions at the same time. To this end, we employ DFPT \cite{baroni_phonons_2001} calculations as implemented in the Quantum Espresso package \cite{giannozzi_quantum_2009} using LDA potentials, a lattice constant of $a=3.24\,$\AA, a vacuum height of $16\,$\AA, a $k$ mesh of $32 \times 32 \times 1$ for the self-consistent electronic calculation, and a $q$ mesh of $8 \times 8 \times 1$ for the phonons. Within the BZ, i.e. for increased $q$ momenta, the most important electron-phonon couplings arise due to acoustic phonon modes in NbS$_2$ (the optical modes couple via a Fröhlich interaction which is proportional to $1/q$ and is thus strongly decreased here). In more detail, the strongest coupling arises due to the LA mode which consequently softens and becomes unstable. 
	
	To estimate an average bare frequency for this mode in the monolayer, we make use of the other acoustic branches which are not at all (ZA) or just slightly (TA) renormalized at the Brillouin zone's M point. Thereby we arrive at an estimation of $\omega_\mathrm{ph} = 20$\,meV for the bare typical phonon frequency, which is comparable to the corresponding modes in bulk NbS$_2$ \cite{leroux_anharmonic_2012}. Using this bare frequency, we re-calculate the renormalized phonon frequency $\omega_\mathrm{ph}^\mathrm{re}(q) = \sqrt{\omega_\mathrm{ph}^2 + 2 \omega_\mathrm{ph} g^2 \chi_0(q)  }$ using the RPA susceptibility $\chi_0(q)$, where we approximate the phonon self-energy as $g^2 \chi_0(q)$. From this, we find instabilities starting from $g_\text{min} \gtrsim 50\,$meV and similar instabilities as in the full DFPT calculation for $g_\mathrm{NbS_2}\approx60\dots70\,$meV. This is comparable to the $g_\mathrm{max}=0.13$\,eV found by Flicker and van Wezel \cite{flicker_charge_2016} for bulk NbSe$_2$. Accordingly, we use an interval of $g = 0.0 \dots 0.1\,$eV in order to study the phonon-induced effects, while $g\approx70\,$meV is supposed to be our material-realistic estimate. At this point it is important to note, that this is clearly just an approximate model. We neglect the phonon dispersions as well as the momentum dependency of the electron-phonon coupling and focus on a single phonon mode only. The latter is, however, well justified since the LA mode is the only mode becoming unstable in DFPT calculations. Furthermore, between the $M$ and $K$ points, the LA mode is rather flat allowing us to describe it as a local Einstein mode. Finally, the Froehlich-like coupling of the those optical modes which have a finite coupling to the electrons is likely underestimated around $\Gamma$ and overestimated around the K point in our model. This means, that a full phonon model might lead to changes in the exact position of the arising divergences in the susceptibilities. In more detail, it is likely that in a full model, the charge instability would emerge more within the Brillouin zone and less at its border.

	\textbf{Dual Boson Approach.} The resulting material-realistic single-band Hubbard-Holstein model is solved using the Dual Boson (DB) method which is based on the Dynamical Mean-Field Theory (DMFT)~\cite{Georges96} philosophy. That is, DB uses an auxiliary single-site problem to take into account strong correlation effects self-consistently. The DB method extends DMFT by also capturing non-local interactions via an effective, dynamic local interaction. Here, as in Ref.~\onlinecite{vanloon_beyond_2014}, the impurity model is determined self-consistently on the Extended Dynamical Mean-Field Theory level. 
	Then, the DB method calculates the momentum and frequency resolved susceptibilities starting with a DMFT-like \textit{interacting} Green's function and then adding non-local vertex corrections (in the ladder approximation) to ensure charge conservation~\cite{Rubtsov12,vanLoon14,Hafermann14-2}. 
	The auxiliary impurity model was solved using a modified version of the open source CT-HYB solver~\cite{Hafermann13,Hafermann14} based on the ALPS libraries~\cite{ALPS2}.
	  
	The Dual Boson calculations use the single-band dispersion $E_\text{mb}(k)$ from the tight-binding model and the effective interaction $U_\mathrm{sb}(q)$ as their input. Both are evaluated on $144 \times 144 \times 1$ $k$ and $q$ meshes. The electron-phonon coupling leads to an additional, retarded, local electron-electron interaction $U^{\text{e-ph}}_{\omega_n}= -2g^2 \frac{\omega_{\text{ph}}}{\omega_{\text{ph}}^2+\omega_n^2}$, where $g$ is the electron-phonon coupling, $\omega_{\text{ph}}$ is the phonon frequency and $\omega_n$ is the $n$-th Matsubara frequency.
 	
	Unless otherwise noted, all Dual Boson simulations were performed at $\beta=25$ eV$^{-1}$ ($T=464\,$K). Calculations without electron-phonon coupling were for temperatures down to $\beta=150$ eV$^{-1}$ ($77\,$K). These showed few qualitative changes: the system remained in the strongly correlated phase. For the case of (strong) electron-phonon coupling, closer to the charge-ordering transition, the temperature is important since ordering is more likely at low temperature, the same holds for spin ordering transitions. However, please note that the model parameters are derived for $T=0\,$K.
	
	To learn more about the role of temperature, we show the local charge and spin susceptibility in Fig.~\ref{fig:localsusc}, the ratio of which is plotted in \ref{fig:AuxSystem}, as a function of $g$ and temperature. Near $g \approx 0.07\,$eV, the low temperature simulations approach the phase transition and the charge susceptibility sees a large change whereas the spin susceptibility develops more smoothly. Comparing the susceptibility with that of the non-interacting system at the same temperature (dashed lines), both temperatures show the same trend, although the magnitude of deviations is generally larger for the low temperature system. At small $g$, $U$ is the dominant interaction and the spin susceptibility is enhanced and the charge susceptibility reduced with respect to the non-interacting system. For both temperatures, we find a coupling strength $g$ where both susceptibilities are enhanced compared to the non-interacting system ($g\approx 0.07\,$eV at the lower temperature and $g\approx 0.85\,$eV at the higher temperature). Thus, we find this simultaneous enhancement of both susceptibilities to be a general feature that does not require a specific temperature. On the other hand, the magnitude and location of the simultaneous enhancement depends on the temperature. In particular, the sharp rise in the charge susceptibility at low temperature signals the approach to the charge-order transition, which depends on temperature.
	
    \section{Data availability statement}	

    The data that support the findings of this study are available from the corresponding author upon reasonable request.		
    
    \section{acknowledgments}
    The authors thank J. van Wezel for useful discussion. 
    E.G.C.P. van Loon and M. R\"osner contributed equally to this project.
    M.R. and G.S. performed the ab-initio determination of the single-band Hamiltonian. E.G.C.P.v.L performed the Dual Boson calculations. All authors contributed to the manuscript.
    M.R. would like to thank the Alexander von Humboldt Foundation for support. E.G.C.P. v. L. and M.I.K. acknowledge support from ERC Advanced Grant 338957 FEMTO/NANO. T. W. and G. S. acknowledge support from DFG via RTG 2247 as well as the European Graphene Flagship.
    
\appendix

	\section*{Supplemental material: Ab initio based parametrization of the Coulomb-interaction model}
	\label{app:CoulombInteraction}
		
		All quantities used in Eq.~(5) of the main text are meant to be given in the same orbital basis as used for the tight-binding model. Furthermore we are dealing here with two-particle quantities which means that $v$, $U$ and $\Pi$ are actually tensors of rank four in the given orbital basis. These tensors can be rearranged to $9 \times 9$ matrices $\mathbf{v}$, $\mathbf{U}$ and $\mathbf{\Pi}$ using a product basis $\tilde{\alpha} = \{ \alpha, \beta \}$ with $\alpha$ and $\beta$ being elements of the single-particle orbital basis. Doing so, Eqs.~(4) and (5) become matrix equations. In order to simplify the following parametrization of the $\mathbf{U}$ matrix we furthermore focus on \emph{density-density} interaction matrix elements only. Thereby the $9 \times 9$ matrices reduce to $3 \times 3$ matrices. 
		
		To describe the $3 \times 3$ matrices of the bare and partially screened Coulomb interaction analytically we make use of the (sorted) eigenbasis of the bare Coulomb interaction $\mathbf{v}$ by diagonalizing it
		\begin{align}
			\mathbf{v}_\text{diag}(q) = 
				\left(
					\begin{array}{ccc}
						v_1(q) & 0 & 0\\
						0 & v_2 & 0 \\
						0 & 0 & v_3
					\end{array} 
				\right),
		\label{eq:BareCoulombInteractionMatrixEigenbasis}
		\end{align}
		where the diagonal matrix elements are given by
		\begin{equation}
			v_i = \braket{ e_i | \mathbf{v} | e_i }
		\end{equation}
		using the eigenvectors of $\mathbf{v}(q \rightarrow 0)$ in their long-wavelength limits
		\begin{equation}
			e_1 = \frac{1}{\sqrt{3}} \begin{pmatrix}  1 \\  1 \\  1 \end{pmatrix}, 
			e_2 = \frac{1}{\sqrt{6}} \begin{pmatrix} +2 \\ -1 \\ -1\end{pmatrix},
			e_3 = \frac{1}{\sqrt{2}} \begin{pmatrix}  0 \\ +1 \\ -1 \end{pmatrix}.
			\label{eq:vEV}
		\end{equation}
		While the leading eigenvalue $v_1(q)$ is a function of $q$ the other two eigenvalues can be readily approximated as constants. The $q$ dependence of the leading eigenvalue can be fitted by
		\begin{equation}
			v_1(q) = \frac{3 e^2}{2 \varepsilon_0 A}  
			         \frac{1}{q(1 + \gamma q + \delta q^2)},
			\label{eq:v1}
		\end{equation}
		where $A$ is the area of the hexagonal unit cell and $\varepsilon_0$ is the vacuum permittivity.
	
		The matrix elements of the partially screened interaction $\mathbf{U}(q)$ in the eigenbasis of the bare interaction $\mathbf{v}(q)$ are then obtained via 
		\begin{equation}
			U_i(q) = \varepsilon_i^{-1}(q) \ v_i(q)
			\label{eq:CoulombInteractionPartiallyScreened}
		\end{equation}
		where $\varepsilon_i(q)$ accounts for the material specific inter-band polarizability due to $\mathbf{\Pi}_\text{rest}(q)$. Its diagonal representation is given by
		\begin{equation}
			\mathbf{\varepsilon}_\text{diag}(q) = 
				\left(
					\begin{array}{ccc}
						\varepsilon_1(q) & 0 & 0\\
						0 & \varepsilon_2 & 0 \\
						0 & 0 & \varepsilon_3 
					\end{array} 
				\right).
		\end{equation}
		\begin{figure}
			\includegraphics[width=0.9\columnwidth]{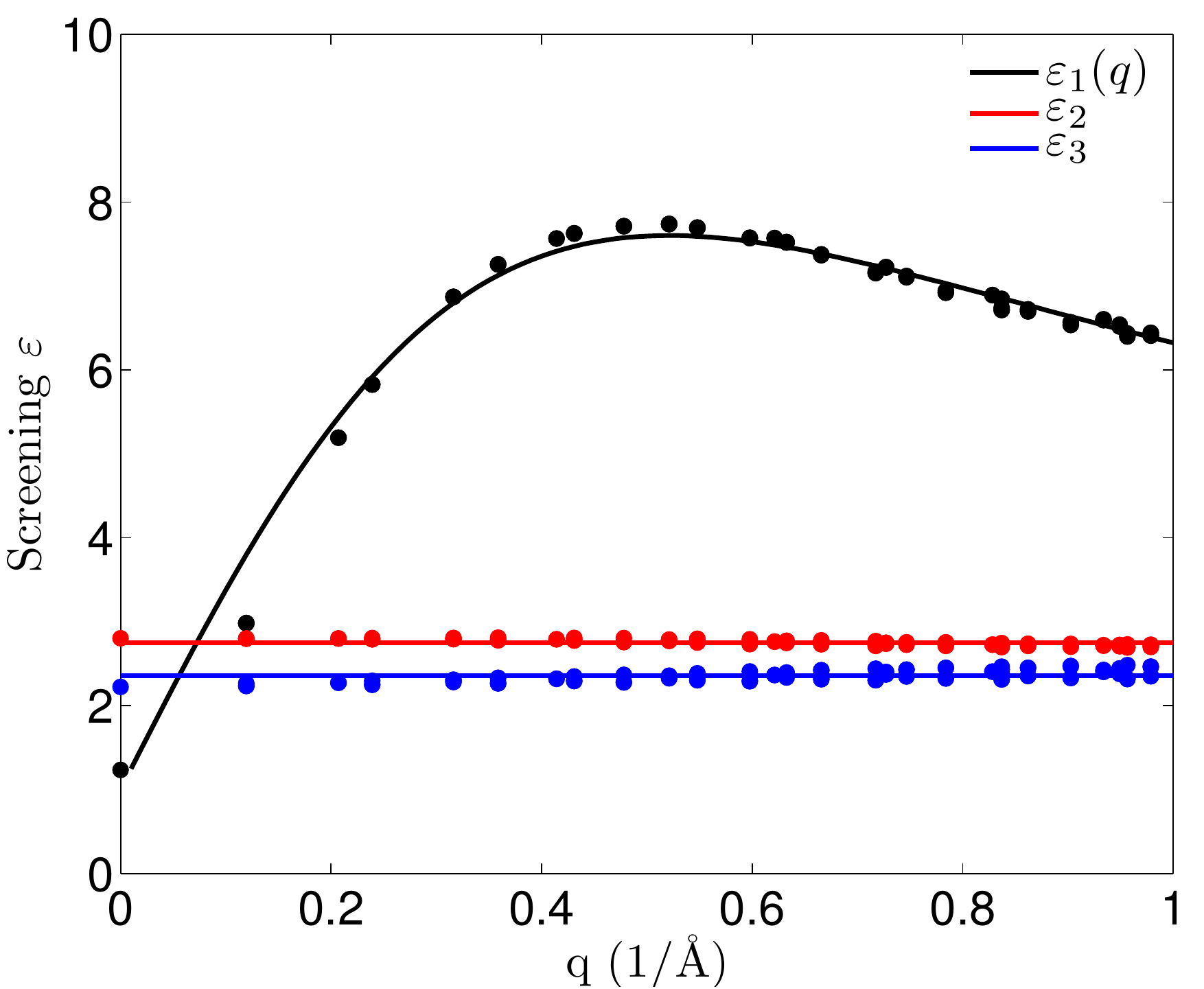}
			\caption{(Color online) Diagonal elements of the dielectric matrix. $\varepsilon_1$ corresponds to the largest (macroscopic) eigenvalue of $\mathbf{v}(q)$ while $\varepsilon_2$ and $\varepsilon_3$ correspond to the microscopic eigenvalues. Circles represent \abinitio data, solid lines show the corresponding analytic fits.}
			\label{fig:EpsilonsMacroMicro}
		\end{figure}%
		Once again, the leading eigenvalue $\varepsilon_1(q)$ is a function of $q$ while the other elements are sufficiently described as constants, as evident from Fig. \ref{fig:EpsilonsMacroMicro}. The former can be expressed by  
		\begin{equation}
			\varepsilon_1(q) = 
				\varepsilon_\infty(q)
				\frac{1 - \beta_1 \beta_2 e^{-2qd}}
				     {1 + (\beta_1 + \beta_2) e^{-qd} + \beta_1 \beta_2 e^{-2qd}}
			\label{eq:BackgroundScreening}
		\end{equation}
		which describes the macroscopic dielectric function of a two-dimensional \emph{semiconductor} (due to the separation of $\Pi(q, \omega)$ as mentioned in the main text, the rest polarization $\Pi_\text{rest}(q)$ does not involve any metallic / intra-band screening anymore) \cite{Keldysh_coulomb_1979}. The parameters $\beta_i$ are defined by
		\begin{equation}
			\beta_i = \frac{\varepsilon_\infty(q) - 1}{\varepsilon_\infty(q) + 1}
		\end{equation}
		and the function $\varepsilon_\infty(q)$ is given by
		\begin{equation}
			\varepsilon_\infty(q) = \frac{a + q^2}
			                             { \frac{a\, \operatorname{sin}(qc)}{qbc} + q^2 } + e.
		\end{equation}
		Thus, we have a closed analytic description of the partially screened Coulomb interaction $\mathbf{U}_\text{diag}(q)$ in the eigenbasis of the bare interaction $\mathbf{v}(q)$ at arbitrary momenta $q$ in the first Brillouin zone. In order to transform it to the original orbital basis we make use of the eigensystem given in Eq. (\ref{eq:vEV}). In fact, this model is appropriate for every two-dimensional transition metal dichalcogenide in its hexagonal structure.

		To fit the model in our case to the data for the NbS$_2$ monolayer we perform full \abinitio\ calculations using the FLEUR and SPEX codes \cite{FLEUR,friedrich_efficient_2009, friedrich_efficient_2010}. Therefore, we start with a DFT calculation for the monolayer using the same parameters as for the tight-binding hopping matrix elements. Afterwards we perform cRPA calculations to get the bare $\mathbf{v}(q)$ and partially screened $\mathbf{U}(q)$ Coulomb interactions in the mentioned Wannier basis. Hereby we exclude the screening arising from the half-filled metallic band. In order to minimize artificial self-screening effects (by periodically repeated images of the NbS$_2$ sheet) we extrapolate the resulting partially screened Coulomb interaction matrix elements from finite super-cell heights of $z = 17$, $22$, $27$, $32$ \AA\ to $z = \infty$ (see Refs. \onlinecite{rosner_wannier_2015,groenewald_valley_2016,schonhoff_interplay_2016} for more details). The resulting interaction matrix elements (in the eigenbasis of the bare interaction) on the given discrete $q$ mesh are finally used to fit all parameters involved in the analytic descriptions from Eqs.~(\ref{eq:BareCoulombInteractionMatrixEigenbasis})--(\ref{eq:BackgroundScreening}). The outcome of this procedure is listed in Tab.~\ref{tab:Fits} and shown in Fig.~\ref{fig:EpsilonsMacroMicro}. 
		 
		\begin{table}[bt]
			\caption{Parameters describing the bare Coulomb interaction $\mathbf{v}(q)$ and the partial static dielectric function $\mathbf{\varepsilon}(q)$ in NbS$_2$.}
			\label{tab:Fits}
			\begin{ruledtabular}
				\begin{tabular}{llll}
				   $\gamma$	& $ +1.923\,$\AA     	& $v_2$				& $0.877\,$eV \\
				   $\delta$	& $ -0.046\,$\AA$^2$	& $v_3$				& $0.393\,$eV \\
				   \hline
				   $a$		& $ +4.657\,$\AA$^{-2}$	& $\varepsilon_2$ 	& $2.749$ \\
				   $b$		& $ +10.53$		      	& $\varepsilon_3$ 	& $2.361$ \\
				   $c$		& $ -0.126\,$\AA	\\
				   $d$		& $ +3.798\,$\AA	\\
				   $e$		& $ +2.610$			\\
				\end{tabular}
			\end{ruledtabular}
		\end{table}
		
		Using these parameters we get a material-realistic description of the $\mathbf{U}(q)$ matrix in the basis of the dominant Nb $d$ orbitals for arbitrary momenta $q$ without the need of redoing any \abinitio calculation.
		
		In combination with the dispersion $E_\text{mb}(k)$ of the half-filled band we thus derived a single-band Hubbard model from the three-band formulation.

\bibliography{bibliography}

\begin{thebibliography}{80}%
\makeatletter
\providecommand \@ifxundefined [1]{%
 \@ifx{#1\undefined}
}%
\providecommand \@ifnum [1]{%
 \ifnum #1\expandafter \@firstoftwo
 \else \expandafter \@secondoftwo
 \fi
}%
\providecommand \@ifx [1]{%
 \ifx #1\expandafter \@firstoftwo
 \else \expandafter \@secondoftwo
 \fi
}%
\providecommand \natexlab [1]{#1}%
\providecommand \enquote  [1]{``#1''}%
\providecommand \bibnamefont  [1]{#1}%
\providecommand \bibfnamefont [1]{#1}%
\providecommand \citenamefont [1]{#1}%
\providecommand \href@noop [0]{\@secondoftwo}%
\providecommand \href [0]{\begingroup \@sanitize@url \@href}%
\providecommand \@href[1]{\@@startlink{#1}\@@href}%
\providecommand \@@href[1]{\endgroup#1\@@endlink}%
\providecommand \@sanitize@url [0]{\catcode `\\12\catcode `\$12\catcode
  `\&12\catcode `\#12\catcode `\^12\catcode `\_12\catcode `\%12\relax}%
\providecommand \@@startlink[1]{}%
\providecommand \@@endlink[0]{}%
\providecommand \url  [0]{\begingroup\@sanitize@url \@url }%
\providecommand \@url [1]{\endgroup\@href {#1}{\urlprefix }}%
\providecommand \urlprefix  [0]{URL }%
\providecommand \Eprint [0]{\href }%
\providecommand \doibase [0]{http://dx.doi.org/}%
\providecommand \selectlanguage [0]{\@gobble}%
\providecommand \bibinfo  [0]{\@secondoftwo}%
\providecommand \bibfield  [0]{\@secondoftwo}%
\providecommand \translation [1]{[#1]}%
\providecommand \BibitemOpen [0]{}%
\providecommand \bibitemStop [0]{}%
\providecommand \bibitemNoStop [0]{.\EOS\space}%
\providecommand \EOS [0]{\spacefactor3000\relax}%
\providecommand \BibitemShut  [1]{\csname bibitem#1\endcsname}%
\let\auto@bib@innerbib\@empty
%</preamble>
\bibitem [{\citenamefont {Yu}\ \emph {et~al.}(2015)\citenamefont {Yu},
  \citenamefont {Yang}, \citenamefont {Lu}, \citenamefont {Yan}, \citenamefont
  {Cho}, \citenamefont {Ma}, \citenamefont {Niu}, \citenamefont {Kim},
  \citenamefont {Son}, \citenamefont {Feng}, \citenamefont {Li}, \citenamefont
  {Cheong}, \citenamefont {Chen},\ and\ \citenamefont
  {Zhang}}]{yu_gate-tunable_2015}%
  \BibitemOpen
  \bibfield  {author} {\bibinfo {author} {\bibfnamefont {Yijun}\ \bibnamefont
  {Yu}}, \bibinfo {author} {\bibfnamefont {Fangyuan}\ \bibnamefont {Yang}},
  \bibinfo {author} {\bibfnamefont {Xiu~Fang}\ \bibnamefont {Lu}}, \bibinfo
  {author} {\bibfnamefont {Ya~Jun}\ \bibnamefont {Yan}}, \bibinfo {author}
  {\bibfnamefont {Yong-Heum}\ \bibnamefont {Cho}}, \bibinfo {author}
  {\bibfnamefont {Liguo}\ \bibnamefont {Ma}}, \bibinfo {author} {\bibfnamefont
  {Xiaohai}\ \bibnamefont {Niu}}, \bibinfo {author} {\bibfnamefont {Sejoong}\
  \bibnamefont {Kim}}, \bibinfo {author} {\bibfnamefont {Young-Woo}\
  \bibnamefont {Son}}, \bibinfo {author} {\bibfnamefont {Donglai}\ \bibnamefont
  {Feng}}, \bibinfo {author} {\bibfnamefont {Shiyan}\ \bibnamefont {Li}},
  \bibinfo {author} {\bibfnamefont {Sang-Wook}\ \bibnamefont {Cheong}},
  \bibinfo {author} {\bibfnamefont {Xian~Hui}\ \bibnamefont {Chen}}, \ and\
  \bibinfo {author} {\bibfnamefont {Yuanbo}\ \bibnamefont {Zhang}},\ }\bibfield
   {title} {\enquote {\bibinfo {title} {Gate-tunable phase transitions in thin
  flakes of {1T-TaS2$_2$}},}\ }\href {\doibase 10.1038/nnano.2014.323}
  {\bibfield  {journal} {\bibinfo  {journal} {Nat. Nano.}\ }\textbf {\bibinfo
  {volume} {10}},\ \bibinfo {pages} {270--276} (\bibinfo {year}
  {2015})}\BibitemShut {NoStop}%
\bibitem [{\citenamefont {Xi}\ \emph {et~al.}(2015)\citenamefont {Xi},
  \citenamefont {Zhao}, \citenamefont {Wang}, \citenamefont {Berger},
  \citenamefont {Forro}, \citenamefont {Shan},\ and\ \citenamefont
  {Mak}}]{xi_strongly_2015}%
  \BibitemOpen
  \bibfield  {author} {\bibinfo {author} {\bibfnamefont {Xiaoxiang}\
  \bibnamefont {Xi}}, \bibinfo {author} {\bibfnamefont {Liang}\ \bibnamefont
  {Zhao}}, \bibinfo {author} {\bibfnamefont {Zefang}\ \bibnamefont {Wang}},
  \bibinfo {author} {\bibfnamefont {Helmuth}\ \bibnamefont {Berger}}, \bibinfo
  {author} {\bibfnamefont {Laszlo}\ \bibnamefont {Forro}}, \bibinfo {author}
  {\bibfnamefont {Jie}\ \bibnamefont {Shan}}, \ and\ \bibinfo {author}
  {\bibfnamefont {Kin~Fai}\ \bibnamefont {Mak}},\ }\bibfield  {title} {\enquote
  {\bibinfo {title} {Strongly enhanced charge-density-wave order in monolayer
  {NbSe}$_2$},}\ }\href {\doibase 10.1038/nnano.2015.143} {\bibfield  {journal}
  {\bibinfo  {journal} {Nat. Nano.}\ }\textbf {\bibinfo {volume} {10}},\
  \bibinfo {pages} {765--769} (\bibinfo {year} {2015})}\BibitemShut {NoStop}%
\bibitem [{\citenamefont {Calandra}\ \emph {et~al.}(2009)\citenamefont
  {Calandra}, \citenamefont {Mazin},\ and\ \citenamefont
  {Mauri}}]{calandra_effect_2009}%
  \BibitemOpen
  \bibfield  {author} {\bibinfo {author} {\bibfnamefont {Matteo}\ \bibnamefont
  {Calandra}}, \bibinfo {author} {\bibfnamefont {I.~I.}\ \bibnamefont {Mazin}},
  \ and\ \bibinfo {author} {\bibfnamefont {Francesco}\ \bibnamefont {Mauri}},\
  }\bibfield  {title} {\enquote {\bibinfo {title} {Effect of dimensionality on
  the charge-density wave in few-layer {2H-NbSe}$_2$},}\ }\href {\doibase
  10.1103/PhysRevB.80.241108} {\bibfield  {journal} {\bibinfo  {journal} {Phys.
  Rev. B}\ }\textbf {\bibinfo {volume} {80}},\ \bibinfo {pages} {241108}
  (\bibinfo {year} {2009})}\BibitemShut {NoStop}%
\bibitem [{\citenamefont {Novoselov}\ \emph {et~al.}(2005)\citenamefont
  {Novoselov}, \citenamefont {Jiang}, \citenamefont {Schedin}, \citenamefont
  {Booth}, \citenamefont {Khotkevich}, \citenamefont {Morozov},\ and\
  \citenamefont {Geim}}]{novoselov_twodimensional_2005}%
  \BibitemOpen
  \bibfield  {author} {\bibinfo {author} {\bibfnamefont {K.~S.}\ \bibnamefont
  {Novoselov}}, \bibinfo {author} {\bibfnamefont {D.}~\bibnamefont {Jiang}},
  \bibinfo {author} {\bibfnamefont {F.}~\bibnamefont {Schedin}}, \bibinfo
  {author} {\bibfnamefont {T.~J.}\ \bibnamefont {Booth}}, \bibinfo {author}
  {\bibfnamefont {V.~V.}\ \bibnamefont {Khotkevich}}, \bibinfo {author}
  {\bibfnamefont {S.~V.}\ \bibnamefont {Morozov}}, \ and\ \bibinfo {author}
  {\bibfnamefont {A.~K.}\ \bibnamefont {Geim}},\ }\bibfield  {title} {\enquote
  {\bibinfo {title} {Two-dimensional atomic crystals},}\ }\href {\doibase
  10.1073/pnas.0502848102} {\bibfield  {journal} {\bibinfo  {journal} {{PNAS}}\
  }\textbf {\bibinfo {volume} {102}},\ \bibinfo {pages} {10451--10453}
  (\bibinfo {year} {2005})}\BibitemShut {NoStop}%
\bibitem [{\citenamefont {Castro~Neto}\ \emph {et~al.}(2009)\citenamefont
  {Castro~Neto}, \citenamefont {Guinea}, \citenamefont {Peres}, \citenamefont
  {Novoselov},\ and\ \citenamefont {Geim}}]{RevModPhys.81.109}%
  \BibitemOpen
  \bibfield  {author} {\bibinfo {author} {\bibfnamefont {A.~H.}\ \bibnamefont
  {Castro~Neto}}, \bibinfo {author} {\bibfnamefont {F.}~\bibnamefont {Guinea}},
  \bibinfo {author} {\bibfnamefont {N.~M.~R.}\ \bibnamefont {Peres}}, \bibinfo
  {author} {\bibfnamefont {K.~S.}\ \bibnamefont {Novoselov}}, \ and\ \bibinfo
  {author} {\bibfnamefont {A.~K.}\ \bibnamefont {Geim}},\ }\bibfield  {title}
  {\enquote {\bibinfo {title} {The electronic properties of graphene},}\ }\href
  {\doibase 10.1103/RevModPhys.81.109} {\bibfield  {journal} {\bibinfo
  {journal} {Rev. Mod. Phys.}\ }\textbf {\bibinfo {volume} {81}},\ \bibinfo
  {pages} {109--162} (\bibinfo {year} {2009})}\BibitemShut {NoStop}%
\bibitem [{\citenamefont {Geim}\ and\ \citenamefont
  {Grigorieva}(2013)}]{geim_van_2013}%
  \BibitemOpen
  \bibfield  {author} {\bibinfo {author} {\bibfnamefont {A.~K.}\ \bibnamefont
  {Geim}}\ and\ \bibinfo {author} {\bibfnamefont {I.~V.}\ \bibnamefont
  {Grigorieva}},\ }\bibfield  {title} {\enquote {\bibinfo {title} {Van der
  {Waals} heterostructures},}\ }\href {\doibase 10.1038/nature12385} {\bibfield
   {journal} {\bibinfo  {journal} {Nature}\ }\textbf {\bibinfo {volume}
  {499}},\ \bibinfo {pages} {419--425} (\bibinfo {year} {2013})}\BibitemShut
  {NoStop}%
\bibitem [{\citenamefont {Animalu}(1964)}]{Animalu_nonlocal_1964}%
  \BibitemOpen
  \bibfield  {author} {\bibinfo {author} {\bibfnamefont {A.~O.E.}\ \bibnamefont
  {Animalu}},\ }\bibfield  {title} {\enquote {\bibinfo {title} {Non-local
  dielectric screening in metals},}\ }\href
  {http://dx.doi.org/10.1080/14786436508221864} {\bibfield  {journal} {\bibinfo
   {journal} {Philosophical Magazine}\ }\textbf {\bibinfo {volume} {11}},\
  \bibinfo {pages} {379--388} (\bibinfo {year} {1964})}\BibitemShut {NoStop}%
\bibitem [{\citenamefont {Keldysh}(1979)}]{Keldysh_coulomb_1979}%
  \BibitemOpen
  \bibfield  {author} {\bibinfo {author} {\bibfnamefont {L.V.}\ \bibnamefont
  {Keldysh}},\ }\bibfield  {title} {\enquote {\bibinfo {title} {Coulomb
  interaction in thin semiconductor and semimetal films},}\ }\href
  {www.jetpletters.ac.ru/ps/1458/article_22207.shtm} {\bibfield  {journal}
  {\bibinfo  {journal} {Pis'ma Zh. Eksp. Teor. Fiz.}\ }\textbf {\bibinfo
  {volume} {39}},\ \bibinfo {pages} {716--719} (\bibinfo {year}
  {1979})}\BibitemShut {NoStop}%
\bibitem [{\citenamefont {Cudazzo}\ \emph {et~al.}(2011)\citenamefont
  {Cudazzo}, \citenamefont {Tokatly},\ and\ \citenamefont
  {Rubio}}]{PhysRevB.84.085406}%
  \BibitemOpen
  \bibfield  {author} {\bibinfo {author} {\bibfnamefont {Pierluigi}\
  \bibnamefont {Cudazzo}}, \bibinfo {author} {\bibfnamefont {Ilya~V.}\
  \bibnamefont {Tokatly}}, \ and\ \bibinfo {author} {\bibfnamefont {Angel}\
  \bibnamefont {Rubio}},\ }\bibfield  {title} {\enquote {\bibinfo {title}
  {Dielectric screening in two-dimensional insulators: Implications for
  excitonic and impurity states in graphane},}\ }\href {\doibase
  10.1103/PhysRevB.84.085406} {\bibfield  {journal} {\bibinfo  {journal} {Phys.
  Rev. B}\ }\textbf {\bibinfo {volume} {84}},\ \bibinfo {pages} {085406}
  (\bibinfo {year} {2011})}\BibitemShut {NoStop}%
\bibitem [{\citenamefont {Andersen}\ \emph {et~al.}(2015)\citenamefont
  {Andersen}, \citenamefont {Latini},\ and\ \citenamefont
  {Thygesen}}]{Kirsten_Dielectric_2015}%
  \BibitemOpen
  \bibfield  {author} {\bibinfo {author} {\bibfnamefont {Kirsten}\ \bibnamefont
  {Andersen}}, \bibinfo {author} {\bibfnamefont {Simone}\ \bibnamefont
  {Latini}}, \ and\ \bibinfo {author} {\bibfnamefont {Kristian~S.}\
  \bibnamefont {Thygesen}},\ }\bibfield  {title} {\enquote {\bibinfo {title}
  {{Dielectric Genome of van der Waals Heterostructures}},}\ }\href {\doibase
  10.1021/acs.nanolett.5b01251} {\bibfield  {journal} {\bibinfo  {journal}
  {Nano Letters}\ }\textbf {\bibinfo {volume} {15}},\ \bibinfo {pages}
  {4616--4621} (\bibinfo {year} {2015})},\ \bibinfo {note} {pMID:
  26047386}\BibitemShut {NoStop}%
\bibitem [{\citenamefont {R{\"o}sner}\ \emph {et~al.}(2015)\citenamefont
  {R{\"o}sner}, \citenamefont {Sasioglu}, \citenamefont {Friedrich},
  \citenamefont {Bl{\"u}gel},\ and\ \citenamefont
  {Wehling}}]{rosner_wannier_2015}%
  \BibitemOpen
  \bibfield  {author} {\bibinfo {author} {\bibfnamefont {M.}~\bibnamefont
  {R{\"o}sner}}, \bibinfo {author} {\bibfnamefont {E.}~\bibnamefont
  {Sasioglu}}, \bibinfo {author} {\bibfnamefont {C.}~\bibnamefont {Friedrich}},
  \bibinfo {author} {\bibfnamefont {S.}~\bibnamefont {Bl{\"u}gel}}, \ and\
  \bibinfo {author} {\bibfnamefont {T.~O.}\ \bibnamefont {Wehling}},\
  }\bibfield  {title} {\enquote {\bibinfo {title} {Wannier function approach to
  realistic {Coulomb} interactions in layered materials and
  heterostructures},}\ }\href {\doibase 10.1103/PhysRevB.92.085102} {\bibfield
  {journal} {\bibinfo  {journal} {Phys. Rev. B}\ }\textbf {\bibinfo {volume}
  {92}},\ \bibinfo {pages} {085102} (\bibinfo {year} {2015})}\BibitemShut
  {NoStop}%
\bibitem [{\citenamefont {Qiu}\ \emph {et~al.}(2016)\citenamefont {Qiu},
  \citenamefont {da~Jornada},\ and\ \citenamefont
  {Louie}}]{PhysRevB.93.235435}%
  \BibitemOpen
  \bibfield  {author} {\bibinfo {author} {\bibfnamefont {Diana~Y.}\
  \bibnamefont {Qiu}}, \bibinfo {author} {\bibfnamefont {Felipe~H.}\
  \bibnamefont {da~Jornada}}, \ and\ \bibinfo {author} {\bibfnamefont
  {Steven~G.}\ \bibnamefont {Louie}},\ }\bibfield  {title} {\enquote {\bibinfo
  {title} {Screening and many-body effects in two-dimensional crystals:
  Monolayer {MoS}$_{2}$},}\ }\href {\doibase 10.1103/PhysRevB.93.235435}
  {\bibfield  {journal} {\bibinfo  {journal} {Phys. Rev. B}\ }\textbf {\bibinfo
  {volume} {93}},\ \bibinfo {pages} {235435} (\bibinfo {year}
  {2016})}\BibitemShut {NoStop}%
\bibitem [{\citenamefont {Nagamatsu}\ \emph {et~al.}(2001)\citenamefont
  {Nagamatsu}, \citenamefont {Nakagawa}, \citenamefont {Muranaka},
  \citenamefont {Zenitani},\ and\ \citenamefont
  {Akimitsu}}]{nagamatsu_superconductivity_2001}%
  \BibitemOpen
  \bibfield  {author} {\bibinfo {author} {\bibfnamefont {J.}~\bibnamefont
  {Nagamatsu}}, \bibinfo {author} {\bibfnamefont {N.}~\bibnamefont {Nakagawa}},
  \bibinfo {author} {\bibfnamefont {T.}~\bibnamefont {Muranaka}}, \bibinfo
  {author} {\bibfnamefont {Y.}~\bibnamefont {Zenitani}}, \ and\ \bibinfo
  {author} {\bibfnamefont {J.}~\bibnamefont {Akimitsu}},\ }\bibfield  {title}
  {\enquote {\bibinfo {title} {Superconductivity at 39 {K} in magnesium
  diboride},}\ }\href {\doibase 10.1038/35065039} {\bibfield  {journal}
  {\bibinfo  {journal} {Nature}\ }\textbf {\bibinfo {volume} {410}},\ \bibinfo
  {pages} {63--64} (\bibinfo {year} {2001})}\BibitemShut {NoStop}%
\bibitem [{\citenamefont {Emery}\ \emph {et~al.}(2005)\citenamefont {Emery},
  \citenamefont {H\'erold}, \citenamefont {d'Astuto}, \citenamefont {Garcia},
  \citenamefont {Bellin}, \citenamefont {Mar\^ech\'e}, \citenamefont
  {Lagrange},\ and\ \citenamefont {Loupias}}]{PhysRevLett.95.087003}%
  \BibitemOpen
  \bibfield  {author} {\bibinfo {author} {\bibfnamefont {N.}~\bibnamefont
  {Emery}}, \bibinfo {author} {\bibfnamefont {C.}~\bibnamefont {H\'erold}},
  \bibinfo {author} {\bibfnamefont {M.}~\bibnamefont {d'Astuto}}, \bibinfo
  {author} {\bibfnamefont {V.}~\bibnamefont {Garcia}}, \bibinfo {author}
  {\bibfnamefont {Ch.}\ \bibnamefont {Bellin}}, \bibinfo {author}
  {\bibfnamefont {J.~F.}\ \bibnamefont {Mar\^ech\'e}}, \bibinfo {author}
  {\bibfnamefont {P.}~\bibnamefont {Lagrange}}, \ and\ \bibinfo {author}
  {\bibfnamefont {G.}~\bibnamefont {Loupias}},\ }\bibfield  {title} {\enquote
  {\bibinfo {title} {Superconductivity of bulk {CaC}$_{6}$},}\ }\href {\doibase
  10.1103/PhysRevLett.95.087003} {\bibfield  {journal} {\bibinfo  {journal}
  {Phys. Rev. Lett.}\ }\textbf {\bibinfo {volume} {95}},\ \bibinfo {pages}
  {087003} (\bibinfo {year} {2005})}\BibitemShut {NoStop}%
\bibitem [{\citenamefont {Ge}\ and\ \citenamefont
  {Liu}(2013)}]{PhysRevB.87.241408}%
  \BibitemOpen
  \bibfield  {author} {\bibinfo {author} {\bibfnamefont {Yizhi}\ \bibnamefont
  {Ge}}\ and\ \bibinfo {author} {\bibfnamefont {Amy~Y.}\ \bibnamefont {Liu}},\
  }\bibfield  {title} {\enquote {\bibinfo {title} {Phonon-mediated
  superconductivity in electron-doped single-layer {MoS}${}_{2}$: A
  first-principles prediction},}\ }\href {\doibase 10.1103/PhysRevB.87.241408}
  {\bibfield  {journal} {\bibinfo  {journal} {Phys. Rev. B}\ }\textbf {\bibinfo
  {volume} {87}},\ \bibinfo {pages} {241408} (\bibinfo {year}
  {2013})}\BibitemShut {NoStop}%
\bibitem [{\citenamefont {R\"osner}\ \emph {et~al.}(2014)\citenamefont
  {R\"osner}, \citenamefont {Haas},\ and\ \citenamefont
  {Wehling}}]{PhysRevB.90.245105}%
  \BibitemOpen
  \bibfield  {author} {\bibinfo {author} {\bibfnamefont {M.}~\bibnamefont
  {R\"osner}}, \bibinfo {author} {\bibfnamefont {S.}~\bibnamefont {Haas}}, \
  and\ \bibinfo {author} {\bibfnamefont {T.~O.}\ \bibnamefont {Wehling}},\
  }\bibfield  {title} {\enquote {\bibinfo {title} {Phase diagram of
  electron-doped dichalcogenides},}\ }\href {\doibase
  10.1103/PhysRevB.90.245105} {\bibfield  {journal} {\bibinfo  {journal} {Phys.
  Rev. B}\ }\textbf {\bibinfo {volume} {90}},\ \bibinfo {pages} {245105}
  (\bibinfo {year} {2014})}\BibitemShut {NoStop}%
\bibitem [{\citenamefont {Wang}\ \emph {et~al.}(2012)\citenamefont {Wang},
  \citenamefont {Kalantar-Zadeh}, \citenamefont {Kis}, \citenamefont
  {Coleman},\ and\ \citenamefont {Strano}}]{wang_electronics_2012}%
  \BibitemOpen
  \bibfield  {author} {\bibinfo {author} {\bibfnamefont {Qing~Hua}\
  \bibnamefont {Wang}}, \bibinfo {author} {\bibfnamefont {Kourosh}\
  \bibnamefont {Kalantar-Zadeh}}, \bibinfo {author} {\bibfnamefont {Andras}\
  \bibnamefont {Kis}}, \bibinfo {author} {\bibfnamefont {Jonathan~N.}\
  \bibnamefont {Coleman}}, \ and\ \bibinfo {author} {\bibfnamefont
  {Michael~S.}\ \bibnamefont {Strano}},\ }\bibfield  {title} {\enquote
  {\bibinfo {title} {Electronics and optoelectronics of two-dimensional
  transition metal dichalcogenides},}\ }\href {\doibase 10.1038/nnano.2012.193}
  {\bibfield  {journal} {\bibinfo  {journal} {Nat Nano}\ }\textbf {\bibinfo
  {volume} {7}},\ \bibinfo {pages} {699--712} (\bibinfo {year}
  {2012})}\BibitemShut {NoStop}%
\bibitem [{\citenamefont {Manzeli}\ \emph {et~al.}(2017)\citenamefont
  {Manzeli}, \citenamefont {Ovchinnikov}, \citenamefont {Pasquier},
  \citenamefont {Yazyev},\ and\ \citenamefont {Kis}}]{manzeli_2d_2017}%
  \BibitemOpen
  \bibfield  {author} {\bibinfo {author} {\bibfnamefont {Sajedeh}\ \bibnamefont
  {Manzeli}}, \bibinfo {author} {\bibfnamefont {Dmitry}\ \bibnamefont
  {Ovchinnikov}}, \bibinfo {author} {\bibfnamefont {Diego}\ \bibnamefont
  {Pasquier}}, \bibinfo {author} {\bibfnamefont {Oleg~V.}\ \bibnamefont
  {Yazyev}}, \ and\ \bibinfo {author} {\bibfnamefont {Andras}\ \bibnamefont
  {Kis}},\ }\bibfield  {title} {\enquote {\bibinfo {title} {2d transition metal
  dichalcogenides},}\ }\href {\doibase 10.1038/natrevmats.2017.33} {\bibfield
  {journal} {\bibinfo  {journal} {Nature Reviews Materials}\ }\textbf {\bibinfo
  {volume} {2}},\ \bibinfo {pages} {natrevmats201733} (\bibinfo {year}
  {2017})}\BibitemShut {NoStop}%
\bibitem [{\citenamefont {Zhuang}\ and\ \citenamefont
  {Hennig}(2016)}]{zhuang_stability_2016}%
  \BibitemOpen
  \bibfield  {author} {\bibinfo {author} {\bibfnamefont {Houlong~L.}\
  \bibnamefont {Zhuang}}\ and\ \bibinfo {author} {\bibfnamefont {Richard~G.}\
  \bibnamefont {Hennig}},\ }\bibfield  {title} {\enquote {\bibinfo {title}
  {Stability and magnetism of strongly correlated single-layer {VS}$_{2}$},}\
  }\href {\doibase 10.1103/PhysRevB.93.054429} {\bibfield  {journal} {\bibinfo
  {journal} {Phys. Rev. B}\ }\textbf {\bibinfo {volume} {93}},\ \bibinfo
  {pages} {054429} (\bibinfo {year} {2016})}\BibitemShut {NoStop}%
\bibitem [{\citenamefont {Isaacs}\ and\ \citenamefont
  {Marianetti}(2016)}]{isaacs_electronic_2016}%
  \BibitemOpen
  \bibfield  {author} {\bibinfo {author} {\bibfnamefont {Eric~B.}\ \bibnamefont
  {Isaacs}}\ and\ \bibinfo {author} {\bibfnamefont {Chris~A.}\ \bibnamefont
  {Marianetti}},\ }\bibfield  {title} {\enquote {\bibinfo {title} {Electronic
  correlations in monolayer {VS}$_{2}$},}\ }\href {\doibase
  10.1103/PhysRevB.94.035120} {\bibfield  {journal} {\bibinfo  {journal} {Phys.
  Rev. B}\ }\textbf {\bibinfo {volume} {94}},\ \bibinfo {pages} {035120}
  (\bibinfo {year} {2016})}\BibitemShut {NoStop}%
\bibitem [{\citenamefont {Mulazzi}\ \emph {et~al.}(2010)\citenamefont
  {Mulazzi}, \citenamefont {Chainani}, \citenamefont {Katayama}, \citenamefont
  {Eguchi}, \citenamefont {Matsunami}, \citenamefont {Ohashi}, \citenamefont
  {Senba}, \citenamefont {Nohara}, \citenamefont {Uchida}, \citenamefont
  {Takagi},\ and\ \citenamefont {Shin}}]{PhysRevB.82.075130}%
  \BibitemOpen
  \bibfield  {author} {\bibinfo {author} {\bibfnamefont {M.}~\bibnamefont
  {Mulazzi}}, \bibinfo {author} {\bibfnamefont {A.}~\bibnamefont {Chainani}},
  \bibinfo {author} {\bibfnamefont {N.}~\bibnamefont {Katayama}}, \bibinfo
  {author} {\bibfnamefont {R.}~\bibnamefont {Eguchi}}, \bibinfo {author}
  {\bibfnamefont {M.}~\bibnamefont {Matsunami}}, \bibinfo {author}
  {\bibfnamefont {H.}~\bibnamefont {Ohashi}}, \bibinfo {author} {\bibfnamefont
  {Y.}~\bibnamefont {Senba}}, \bibinfo {author} {\bibfnamefont
  {M.}~\bibnamefont {Nohara}}, \bibinfo {author} {\bibfnamefont
  {M.}~\bibnamefont {Uchida}}, \bibinfo {author} {\bibfnamefont
  {H.}~\bibnamefont {Takagi}}, \ and\ \bibinfo {author} {\bibfnamefont
  {S.}~\bibnamefont {Shin}},\ }\bibfield  {title} {\enquote {\bibinfo {title}
  {Absence of nesting in the charge-density-wave system {1T-VS}$_{2}$ as seen
  by photoelectron spectroscopy},}\ }\href {\doibase
  10.1103/PhysRevB.82.075130} {\bibfield  {journal} {\bibinfo  {journal} {Phys.
  Rev. B}\ }\textbf {\bibinfo {volume} {82}},\ \bibinfo {pages} {075130}
  (\bibinfo {year} {2010})}\BibitemShut {NoStop}%
\bibitem [{\citenamefont {Sun}\ \emph {et~al.}(2015)\citenamefont {Sun},
  \citenamefont {Yao}, \citenamefont {Hu}, \citenamefont {Guo}, \citenamefont
  {Liu}, \citenamefont {Wei},\ and\ \citenamefont {Wu}}]{C5CP01326G}%
  \BibitemOpen
  \bibfield  {author} {\bibinfo {author} {\bibfnamefont {Xu}~\bibnamefont
  {Sun}}, \bibinfo {author} {\bibfnamefont {Tao}\ \bibnamefont {Yao}}, \bibinfo
  {author} {\bibfnamefont {Zhenpeng}\ \bibnamefont {Hu}}, \bibinfo {author}
  {\bibfnamefont {Yuqiao}\ \bibnamefont {Guo}}, \bibinfo {author}
  {\bibfnamefont {Qinghua}\ \bibnamefont {Liu}}, \bibinfo {author}
  {\bibfnamefont {Shiqiang}\ \bibnamefont {Wei}}, \ and\ \bibinfo {author}
  {\bibfnamefont {Changzheng}\ \bibnamefont {Wu}},\ }\bibfield  {title}
  {\enquote {\bibinfo {title} {In situ unravelling structural modulation across
  the charge-density-wave transition in vanadium disulfide},}\ }\href {\doibase
  10.1039/C5CP01326G} {\bibfield  {journal} {\bibinfo  {journal} {Phys. Chem.
  Chem. Phys.}\ }\textbf {\bibinfo {volume} {17}},\ \bibinfo {pages}
  {13333--13339} (\bibinfo {year} {2015})}\BibitemShut {NoStop}%
\bibitem [{\citenamefont {Gauzzi}\ \emph {et~al.}(2014)\citenamefont {Gauzzi},
  \citenamefont {Sellam}, \citenamefont {Rousse}, \citenamefont {Klein},
  \citenamefont {Taverna}, \citenamefont {Giura}, \citenamefont {Calandra},
  \citenamefont {Loupias}, \citenamefont {Gozzo}, \citenamefont {Gilioli},
  \citenamefont {Bolzoni}, \citenamefont {Allodi}, \citenamefont {De~Renzi},
  \citenamefont {Calestani},\ and\ \citenamefont {Roy}}]{PhysRevB.89.235125}%
  \BibitemOpen
  \bibfield  {author} {\bibinfo {author} {\bibfnamefont {A.}~\bibnamefont
  {Gauzzi}}, \bibinfo {author} {\bibfnamefont {A.}~\bibnamefont {Sellam}},
  \bibinfo {author} {\bibfnamefont {G.}~\bibnamefont {Rousse}}, \bibinfo
  {author} {\bibfnamefont {Y.}~\bibnamefont {Klein}}, \bibinfo {author}
  {\bibfnamefont {D.}~\bibnamefont {Taverna}}, \bibinfo {author} {\bibfnamefont
  {P.}~\bibnamefont {Giura}}, \bibinfo {author} {\bibfnamefont
  {M.}~\bibnamefont {Calandra}}, \bibinfo {author} {\bibfnamefont
  {G.}~\bibnamefont {Loupias}}, \bibinfo {author} {\bibfnamefont
  {F.}~\bibnamefont {Gozzo}}, \bibinfo {author} {\bibfnamefont
  {E.}~\bibnamefont {Gilioli}}, \bibinfo {author} {\bibfnamefont
  {F.}~\bibnamefont {Bolzoni}}, \bibinfo {author} {\bibfnamefont
  {G.}~\bibnamefont {Allodi}}, \bibinfo {author} {\bibfnamefont
  {R.}~\bibnamefont {De~Renzi}}, \bibinfo {author} {\bibfnamefont {G.~L.}\
  \bibnamefont {Calestani}}, \ and\ \bibinfo {author} {\bibfnamefont
  {P.}~\bibnamefont {Roy}},\ }\bibfield  {title} {\enquote {\bibinfo {title}
  {Possible phase separation and weak localization in the absence of a
  charge-density wave in single-phase {1T-VS}$_{2}$},}\ }\href {\doibase
  10.1103/PhysRevB.89.235125} {\bibfield  {journal} {\bibinfo  {journal} {Phys.
  Rev. B}\ }\textbf {\bibinfo {volume} {89}},\ \bibinfo {pages} {235125}
  (\bibinfo {year} {2014})}\BibitemShut {NoStop}%
\bibitem [{\citenamefont {Xu}\ \emph {et~al.}(2013)\citenamefont {Xu},
  \citenamefont {Chen}, \citenamefont {Li}, \citenamefont {Wu}, \citenamefont
  {Guo}, \citenamefont {Zhao}, \citenamefont {Wu},\ and\ \citenamefont
  {Xie}}]{ANIE:ANIE201304337}%
  \BibitemOpen
  \bibfield  {author} {\bibinfo {author} {\bibfnamefont {Kun}\ \bibnamefont
  {Xu}}, \bibinfo {author} {\bibfnamefont {Pengzuo}\ \bibnamefont {Chen}},
  \bibinfo {author} {\bibfnamefont {Xiuling}\ \bibnamefont {Li}}, \bibinfo
  {author} {\bibfnamefont {Changzheng}\ \bibnamefont {Wu}}, \bibinfo {author}
  {\bibfnamefont {Yuqiao}\ \bibnamefont {Guo}}, \bibinfo {author}
  {\bibfnamefont {Jiyin}\ \bibnamefont {Zhao}}, \bibinfo {author}
  {\bibfnamefont {Xiaojun}\ \bibnamefont {Wu}}, \ and\ \bibinfo {author}
  {\bibfnamefont {Yi}~\bibnamefont {Xie}},\ }\bibfield  {title} {\enquote
  {\bibinfo {title} {Ultrathin nanosheets of vanadium diselenide: A metallic
  two-dimensional material with ferromagnetic charge-density-wave behavior},}\
  }\href {\doibase 10.1002/anie.201304337} {\bibfield  {journal} {\bibinfo
  {journal} {Angewandte Chemie International Edition}\ }\textbf {\bibinfo
  {volume} {52}},\ \bibinfo {pages} {10477--10481} (\bibinfo {year}
  {2013})}\BibitemShut {NoStop}%
\bibitem [{\citenamefont {Pillo}\ \emph {et~al.}(2000)\citenamefont {Pillo},
  \citenamefont {Hayoz}, \citenamefont {Berger}, \citenamefont {Fasel},
  \citenamefont {Schlapbach},\ and\ \citenamefont
  {Aebi}}]{pillo_interplay_2000}%
  \BibitemOpen
  \bibfield  {author} {\bibinfo {author} {\bibfnamefont {Th.}\ \bibnamefont
  {Pillo}}, \bibinfo {author} {\bibfnamefont {J.}~\bibnamefont {Hayoz}},
  \bibinfo {author} {\bibfnamefont {H.}~\bibnamefont {Berger}}, \bibinfo
  {author} {\bibfnamefont {R.}~\bibnamefont {Fasel}}, \bibinfo {author}
  {\bibfnamefont {L.}~\bibnamefont {Schlapbach}}, \ and\ \bibinfo {author}
  {\bibfnamefont {P.}~\bibnamefont {Aebi}},\ }\bibfield  {title} {\enquote
  {\bibinfo {title} {Interplay between electron-electron interaction and
  electron-phonon coupling near the fermi surface of {1T-TaS}$_{2}$},}\ }\href
  {\doibase 10.1103/PhysRevB.62.4277} {\bibfield  {journal} {\bibinfo
  {journal} {Phys. Rev. B}\ }\textbf {\bibinfo {volume} {62}},\ \bibinfo
  {pages} {4277--4287} (\bibinfo {year} {2000})}\BibitemShut {NoStop}%
\bibitem [{\citenamefont {Cho}\ \emph {et~al.}(2015)\citenamefont {Cho},
  \citenamefont {Cho}, \citenamefont {Cheong}, \citenamefont {Kim},\ and\
  \citenamefont {Yeom}}]{cho_interplay_2015}%
  \BibitemOpen
  \bibfield  {author} {\bibinfo {author} {\bibfnamefont {Doohee}\ \bibnamefont
  {Cho}}, \bibinfo {author} {\bibfnamefont {Yong-Heum}\ \bibnamefont {Cho}},
  \bibinfo {author} {\bibfnamefont {Sang-Wook}\ \bibnamefont {Cheong}},
  \bibinfo {author} {\bibfnamefont {Ki-Seok}\ \bibnamefont {Kim}}, \ and\
  \bibinfo {author} {\bibfnamefont {Han~Woong}\ \bibnamefont {Yeom}},\
  }\bibfield  {title} {\enquote {\bibinfo {title} {Interplay of
  electron-electron and electron-phonon interactions in the low-temperature
  phase of {1T-TaS}$_{2}$},}\ }\href {\doibase 10.1103/PhysRevB.92.085132}
  {\bibfield  {journal} {\bibinfo  {journal} {Phys. Rev. B}\ }\textbf {\bibinfo
  {volume} {92}},\ \bibinfo {pages} {085132} (\bibinfo {year}
  {2015})}\BibitemShut {NoStop}%
\bibitem [{\citenamefont {Cho}\ \emph {et~al.}(2016)\citenamefont {Cho},
  \citenamefont {Cheon}, \citenamefont {Kim}, \citenamefont {Lee},
  \citenamefont {Cho}, \citenamefont {Cheong},\ and\ \citenamefont
  {Yeom}}]{cho_nanoscale_2016}%
  \BibitemOpen
  \bibfield  {author} {\bibinfo {author} {\bibfnamefont {Doohee}\ \bibnamefont
  {Cho}}, \bibinfo {author} {\bibfnamefont {Sangmo}\ \bibnamefont {Cheon}},
  \bibinfo {author} {\bibfnamefont {Ki-Seok}\ \bibnamefont {Kim}}, \bibinfo
  {author} {\bibfnamefont {Sung-Hoon}\ \bibnamefont {Lee}}, \bibinfo {author}
  {\bibfnamefont {Yong-Heum}\ \bibnamefont {Cho}}, \bibinfo {author}
  {\bibfnamefont {Sang-Wook}\ \bibnamefont {Cheong}}, \ and\ \bibinfo {author}
  {\bibfnamefont {Han~Woong}\ \bibnamefont {Yeom}},\ }\bibfield  {title}
  {\enquote {\bibinfo {title} {Nanoscale manipulation of the {Mott} insulating
  state coupled to charge order in {1T-TaS}$_{2}$},}\ }\href {\doibase
  10.1038/ncomms10453} {\bibfield  {journal} {\bibinfo  {journal} {Nature
  Communications}\ }\textbf {\bibinfo {volume} {7}},\ \bibinfo {pages}
  {ncomms10453} (\bibinfo {year} {2016})}\BibitemShut {NoStop}%
\bibitem [{\citenamefont {Ma}\ \emph {et~al.}(2016)\citenamefont {Ma},
  \citenamefont {Ye}, \citenamefont {Yu}, \citenamefont {Lu}, \citenamefont
  {Niu}, \citenamefont {Kim}, \citenamefont {Feng}, \citenamefont {Tománek},
  \citenamefont {Son}, \citenamefont {Chen},\ and\ \citenamefont
  {Zhang}}]{ma_metallic_2016}%
  \BibitemOpen
  \bibfield  {author} {\bibinfo {author} {\bibfnamefont {Liguo}\ \bibnamefont
  {Ma}}, \bibinfo {author} {\bibfnamefont {Cun}\ \bibnamefont {Ye}}, \bibinfo
  {author} {\bibfnamefont {Yijun}\ \bibnamefont {Yu}}, \bibinfo {author}
  {\bibfnamefont {Xiu~Fang}\ \bibnamefont {Lu}}, \bibinfo {author}
  {\bibfnamefont {Xiaohai}\ \bibnamefont {Niu}}, \bibinfo {author}
  {\bibfnamefont {Sejoong}\ \bibnamefont {Kim}}, \bibinfo {author}
  {\bibfnamefont {Donglai}\ \bibnamefont {Feng}}, \bibinfo {author}
  {\bibfnamefont {David}\ \bibnamefont {Tománek}}, \bibinfo {author}
  {\bibfnamefont {Young-Woo}\ \bibnamefont {Son}}, \bibinfo {author}
  {\bibfnamefont {Xian~Hui}\ \bibnamefont {Chen}}, \ and\ \bibinfo {author}
  {\bibfnamefont {Yuanbo}\ \bibnamefont {Zhang}},\ }\bibfield  {title}
  {\enquote {\bibinfo {title} {A metallic mosaic phase and the origin of
  {Mott}-insulating state in 1t-{TaS}$_{\textrm{2}}$},}\ }\href {\doibase
  10.1038/ncomms10956} {\bibfield  {journal} {\bibinfo  {journal} {Nature
  Communications}\ }\textbf {\bibinfo {volume} {7}},\ \bibinfo {pages}
  {ncomms10956} (\bibinfo {year} {2016})}\BibitemShut {NoStop}%
\bibitem [{\citenamefont {{Sipos B.}}\ \emph {et~al.}(2008)\citenamefont
  {{Sipos B.}}, \citenamefont {{Kusmartseva A. F.}}, \citenamefont {{Akrap
  A.}}, \citenamefont {{Berger H.}}, \citenamefont {{Forr{\'o} L.}},\ and\
  \citenamefont {{Tuti\v{s} E.}}}]{sipos_mott_2008}%
  \BibitemOpen
  \bibfield  {author} {\bibinfo {author} {\bibnamefont {{Sipos B.}}}, \bibinfo
  {author} {\bibnamefont {{Kusmartseva A. F.}}}, \bibinfo {author}
  {\bibnamefont {{Akrap A.}}}, \bibinfo {author} {\bibnamefont {{Berger H.}}},
  \bibinfo {author} {\bibnamefont {{Forr{\'o} L.}}}, \ and\ \bibinfo {author}
  {\bibnamefont {{Tuti\v{s} E.}}},\ }\bibfield  {title} {\enquote {\bibinfo
  {title} {{From Mott state to superconductivity in 1T-TaS2}},}\ }\href
  {\doibase http://dx.doi.org/10.1038/nmat2318 10.1038/nmat2318} {\bibfield
  {journal} {\bibinfo  {journal} {Nature Materials}\ }\textbf {\bibinfo
  {volume} {7}},\ \bibinfo {pages} {960} (\bibinfo {year} {2008})}\BibitemShut
  {NoStop}%
\bibitem [{\citenamefont {Liu}(2009)}]{liu_electron-phonon_2009}%
  \BibitemOpen
  \bibfield  {author} {\bibinfo {author} {\bibfnamefont {Amy~Y.}\ \bibnamefont
  {Liu}},\ }\bibfield  {title} {\enquote {\bibinfo {title} {Electron-phonon
  coupling in compressed {1T-TaS}$_{2}$: Stability and superconductivity from
  first principles},}\ }\href {\doibase 10.1103/PhysRevB.79.220515} {\bibfield
  {journal} {\bibinfo  {journal} {Phys. Rev. B}\ }\textbf {\bibinfo {volume}
  {79}},\ \bibinfo {pages} {220515} (\bibinfo {year} {2009})}\BibitemShut
  {NoStop}%
\bibitem [{\citenamefont {Leroux}\ \emph {et~al.}(2015)\citenamefont {Leroux},
  \citenamefont {Errea}, \citenamefont {Le~Tacon}, \citenamefont {Souliou},
  \citenamefont {Garbarino}, \citenamefont {Cario}, \citenamefont {Bosak},
  \citenamefont {Mauri}, \citenamefont {Calandra},\ and\ \citenamefont
  {Rodi\`ere}}]{PhysRevB.92.140303}%
  \BibitemOpen
  \bibfield  {author} {\bibinfo {author} {\bibfnamefont {Maxime}\ \bibnamefont
  {Leroux}}, \bibinfo {author} {\bibfnamefont {Ion}\ \bibnamefont {Errea}},
  \bibinfo {author} {\bibfnamefont {Mathieu}\ \bibnamefont {Le~Tacon}},
  \bibinfo {author} {\bibfnamefont {Sofia-Michaela}\ \bibnamefont {Souliou}},
  \bibinfo {author} {\bibfnamefont {Gaston}\ \bibnamefont {Garbarino}},
  \bibinfo {author} {\bibfnamefont {Laurent}\ \bibnamefont {Cario}}, \bibinfo
  {author} {\bibfnamefont {Alexey}\ \bibnamefont {Bosak}}, \bibinfo {author}
  {\bibfnamefont {Francesco}\ \bibnamefont {Mauri}}, \bibinfo {author}
  {\bibfnamefont {Matteo}\ \bibnamefont {Calandra}}, \ and\ \bibinfo {author}
  {\bibfnamefont {Pierre}\ \bibnamefont {Rodi\`ere}},\ }\bibfield  {title}
  {\enquote {\bibinfo {title} {Strong anharmonicity induces quantum melting of
  charge density wave in {2H-NbSe}$_{2}$ under pressure},}\ }\href {\doibase
  10.1103/PhysRevB.92.140303} {\bibfield  {journal} {\bibinfo  {journal} {Phys.
  Rev. B}\ }\textbf {\bibinfo {volume} {92}},\ \bibinfo {pages} {140303}
  (\bibinfo {year} {2015})}\BibitemShut {NoStop}%
\bibitem [{\citenamefont {Leroux}\ \emph {et~al.}(2012)\citenamefont {Leroux},
  \citenamefont {Le~Tacon}, \citenamefont {Calandra}, \citenamefont {Cario},
  \citenamefont {Measson}, \citenamefont {Diener}, \citenamefont {Borrissenko},
  \citenamefont {Bosak},\ and\ \citenamefont
  {Rodiere}}]{leroux_anharmonic_2012}%
  \BibitemOpen
  \bibfield  {author} {\bibinfo {author} {\bibfnamefont {M.}~\bibnamefont
  {Leroux}}, \bibinfo {author} {\bibfnamefont {M.}~\bibnamefont {Le~Tacon}},
  \bibinfo {author} {\bibfnamefont {M.}~\bibnamefont {Calandra}}, \bibinfo
  {author} {\bibfnamefont {L.}~\bibnamefont {Cario}}, \bibinfo {author}
  {\bibfnamefont {M-A.}\ \bibnamefont {Measson}}, \bibinfo {author}
  {\bibfnamefont {P.}~\bibnamefont {Diener}}, \bibinfo {author} {\bibfnamefont
  {E.}~\bibnamefont {Borrissenko}}, \bibinfo {author} {\bibfnamefont
  {A.}~\bibnamefont {Bosak}}, \ and\ \bibinfo {author} {\bibfnamefont
  {P.}~\bibnamefont {Rodiere}},\ }\bibfield  {title} {\enquote {\bibinfo
  {title} {Anharmonic suppression of charge density waves in {2H-NbS}$_2$},}\
  }\href {\doibase 10.1103/PhysRevB.86.155125} {\bibfield  {journal} {\bibinfo
  {journal} {Phys. Rev. B}\ }\textbf {\bibinfo {volume} {86}},\ \bibinfo
  {pages} {155125} (\bibinfo {year} {2012})}\BibitemShut {NoStop}%
\bibitem [{\citenamefont {{Flicker Felix}}\ and\ \citenamefont {van
  Wezel~Jasper}(2015)}]{flicker_charge_2015}%
  \BibitemOpen
  \bibfield  {author} {\bibinfo {author} {\bibnamefont {{Flicker Felix}}}\ and\
  \bibinfo {author} {\bibnamefont {van Wezel~Jasper}},\ }\bibfield  {title}
  {\enquote {\bibinfo {title} {{Charge order from orbital-dependent coupling
  evidenced by NbSe2}},}\ }\href {\doibase http://dx.doi.org/10.1038/ncomms8034
  10.1038/ncomms8034} {\bibfield  {journal} {\bibinfo  {journal} {Nature
  Communications}\ }\textbf {\bibinfo {volume} {6}},\ \bibinfo {pages} {7034}
  (\bibinfo {year} {2015})}\BibitemShut {NoStop}%
\bibitem [{\citenamefont {Flicker}\ and\ \citenamefont {van
  Wezel}(2015)}]{flicker_charge_2015-1}%
  \BibitemOpen
  \bibfield  {author} {\bibinfo {author} {\bibfnamefont {Felix}\ \bibnamefont
  {Flicker}}\ and\ \bibinfo {author} {\bibfnamefont {Jasper}\ \bibnamefont {van
  Wezel}},\ }\bibfield  {title} {\enquote {\bibinfo {title} {Charge ordering
  geometries in uniaxially strained {NbSe}$_2$},}\ }\href {\doibase
  10.1103/PhysRevB.92.201103} {\bibfield  {journal} {\bibinfo  {journal} {Phys.
  Rev. B}\ }\textbf {\bibinfo {volume} {92}},\ \bibinfo {pages} {201103}
  (\bibinfo {year} {2015})}\BibitemShut {NoStop}%
\bibitem [{\citenamefont {Flicker}\ and\ \citenamefont {van
  Wezel}(2016)}]{flicker_charge_2016}%
  \BibitemOpen
  \bibfield  {author} {\bibinfo {author} {\bibfnamefont {Felix}\ \bibnamefont
  {Flicker}}\ and\ \bibinfo {author} {\bibfnamefont {Jasper}\ \bibnamefont {van
  Wezel}},\ }\bibfield  {title} {\enquote {\bibinfo {title} {Charge order in
  {NbSe}$_{2}$},}\ }\href {\doibase 10.1103/PhysRevB.94.235135} {\bibfield
  {journal} {\bibinfo  {journal} {Phys. Rev. B}\ }\textbf {\bibinfo {volume}
  {94}},\ \bibinfo {pages} {235135} (\bibinfo {year} {2016})}\BibitemShut
  {NoStop}%
\bibitem [{\citenamefont {Chatterjee}\ \emph {et~al.}(2015)\citenamefont
  {Chatterjee}, \citenamefont {Zhao}, \citenamefont {Iavarone}, \citenamefont
  {Capua}, \citenamefont {Castellan}, \citenamefont {Karapetrov}, \citenamefont
  {Malliakas}, \citenamefont {Kanatzidis}, \citenamefont {Claus}, \citenamefont
  {Ruff}, \citenamefont {Weber}, \citenamefont {Wezel}, \citenamefont
  {Campuzano}, \citenamefont {Osborn}, \citenamefont {Randeria}, \citenamefont
  {Trivedi}, \citenamefont {Norman},\ and\ \citenamefont
  {Rosenkranz}}]{chatterjee_emergence_2015}%
  \BibitemOpen
  \bibfield  {author} {\bibinfo {author} {\bibfnamefont {U.}~\bibnamefont
  {Chatterjee}}, \bibinfo {author} {\bibfnamefont {J.}~\bibnamefont {Zhao}},
  \bibinfo {author} {\bibfnamefont {M.}~\bibnamefont {Iavarone}}, \bibinfo
  {author} {\bibfnamefont {R.~Di}\ \bibnamefont {Capua}}, \bibinfo {author}
  {\bibfnamefont {J.~P.}\ \bibnamefont {Castellan}}, \bibinfo {author}
  {\bibfnamefont {G.}~\bibnamefont {Karapetrov}}, \bibinfo {author}
  {\bibfnamefont {C.~D.}\ \bibnamefont {Malliakas}}, \bibinfo {author}
  {\bibfnamefont {M.~G.}\ \bibnamefont {Kanatzidis}}, \bibinfo {author}
  {\bibfnamefont {H.}~\bibnamefont {Claus}}, \bibinfo {author} {\bibfnamefont
  {J.~P.~C.}\ \bibnamefont {Ruff}}, \bibinfo {author} {\bibfnamefont
  {F.}~\bibnamefont {Weber}}, \bibinfo {author} {\bibfnamefont {J.~van}\
  \bibnamefont {Wezel}}, \bibinfo {author} {\bibfnamefont {J.~C.}\ \bibnamefont
  {Campuzano}}, \bibinfo {author} {\bibfnamefont {R.}~\bibnamefont {Osborn}},
  \bibinfo {author} {\bibfnamefont {M.}~\bibnamefont {Randeria}}, \bibinfo
  {author} {\bibfnamefont {N.}~\bibnamefont {Trivedi}}, \bibinfo {author}
  {\bibfnamefont {M.~R.}\ \bibnamefont {Norman}}, \ and\ \bibinfo {author}
  {\bibfnamefont {S.}~\bibnamefont {Rosenkranz}},\ }\bibfield  {title}
  {\enquote {\bibinfo {title} {Emergence of coherence in the charge-density
  wave state of 2\textit{{H}}-{NbSe}$_{\textrm{2}}$},}\ }\href {\doibase
  10.1038/ncomms7313} {\bibfield  {journal} {\bibinfo  {journal} {Nature
  Communications}\ }\textbf {\bibinfo {volume} {6}},\ \bibinfo {pages}
  {ncomms7313} (\bibinfo {year} {2015})}\BibitemShut {NoStop}%
\bibitem [{\citenamefont {Ugeda}\ \emph {et~al.}(2016)\citenamefont {Ugeda},
  \citenamefont {Bradley}, \citenamefont {Zhang}, \citenamefont {Onishi},
  \citenamefont {Chen}, \citenamefont {Ruan}, \citenamefont
  {Ojeda-Aristizabal}, \citenamefont {Ryu}, \citenamefont {Edmonds},
  \citenamefont {Tsai}, \citenamefont {Riss}, \citenamefont {Mo}, \citenamefont
  {Lee}, \citenamefont {Zettl}, \citenamefont {Hussain}, \citenamefont {Shen},\
  and\ \citenamefont {Crommie}}]{ugeda_characterization_2016}%
  \BibitemOpen
  \bibfield  {author} {\bibinfo {author} {\bibfnamefont {Miguel~M.}\
  \bibnamefont {Ugeda}}, \bibinfo {author} {\bibfnamefont {Aaron~J.}\
  \bibnamefont {Bradley}}, \bibinfo {author} {\bibfnamefont {Yi}~\bibnamefont
  {Zhang}}, \bibinfo {author} {\bibfnamefont {Seita}\ \bibnamefont {Onishi}},
  \bibinfo {author} {\bibfnamefont {Yi}~\bibnamefont {Chen}}, \bibinfo {author}
  {\bibfnamefont {Wei}\ \bibnamefont {Ruan}}, \bibinfo {author} {\bibfnamefont
  {Claudia}\ \bibnamefont {Ojeda-Aristizabal}}, \bibinfo {author}
  {\bibfnamefont {Hyejin}\ \bibnamefont {Ryu}}, \bibinfo {author}
  {\bibfnamefont {Mark~T.}\ \bibnamefont {Edmonds}}, \bibinfo {author}
  {\bibfnamefont {Hsin-Zon}\ \bibnamefont {Tsai}}, \bibinfo {author}
  {\bibfnamefont {Alexander}\ \bibnamefont {Riss}}, \bibinfo {author}
  {\bibfnamefont {Sung-Kwan}\ \bibnamefont {Mo}}, \bibinfo {author}
  {\bibfnamefont {Dunghai}\ \bibnamefont {Lee}}, \bibinfo {author}
  {\bibfnamefont {Alex}\ \bibnamefont {Zettl}}, \bibinfo {author}
  {\bibfnamefont {Zahid}\ \bibnamefont {Hussain}}, \bibinfo {author}
  {\bibfnamefont {Zhi-Xun}\ \bibnamefont {Shen}}, \ and\ \bibinfo {author}
  {\bibfnamefont {Michael~F.}\ \bibnamefont {Crommie}},\ }\bibfield  {title}
  {\enquote {\bibinfo {title} {Characterization of collective ground states in
  single-layer {NbSe}2},}\ }\href {\doibase 10.1038/nphys3527} {\bibfield
  {journal} {\bibinfo  {journal} {Nat. Phys.}\ }\textbf {\bibinfo {volume}
  {12}},\ \bibinfo {pages} {92--97} (\bibinfo {year} {2016})}\BibitemShut
  {NoStop}%
\bibitem [{\citenamefont {Nakata}\ \emph {et~al.}(2016)\citenamefont {Nakata},
  \citenamefont {Sugawara}, \citenamefont {Shimizu}, \citenamefont {Okada},
  \citenamefont {Han}, \citenamefont {Hitosugi}, \citenamefont {Ueno},
  \citenamefont {Sato},\ and\ \citenamefont
  {Takahashi}}]{nakata_monolayer_2016}%
  \BibitemOpen
  \bibfield  {author} {\bibinfo {author} {\bibfnamefont {Yuki}\ \bibnamefont
  {Nakata}}, \bibinfo {author} {\bibfnamefont {Katsuaki}\ \bibnamefont
  {Sugawara}}, \bibinfo {author} {\bibfnamefont {Ryota}\ \bibnamefont
  {Shimizu}}, \bibinfo {author} {\bibfnamefont {Yoshinori}\ \bibnamefont
  {Okada}}, \bibinfo {author} {\bibfnamefont {Patrick}\ \bibnamefont {Han}},
  \bibinfo {author} {\bibfnamefont {Taro}\ \bibnamefont {Hitosugi}}, \bibinfo
  {author} {\bibfnamefont {Keiji}\ \bibnamefont {Ueno}}, \bibinfo {author}
  {\bibfnamefont {Takafumi}\ \bibnamefont {Sato}}, \ and\ \bibinfo {author}
  {\bibfnamefont {Takashi}\ \bibnamefont {Takahashi}},\ }\bibfield  {title}
  {\enquote {\bibinfo {title} {Monolayer {1T}-{NbSe}$_2$ as a {Mott}
  insulator},}\ }\href {\doibase 10.1038/am.2016.157} {\bibfield  {journal}
  {\bibinfo  {journal} {NPG Asia Mater}\ }\textbf {\bibinfo {volume} {8}},\
  \bibinfo {pages} {e321} (\bibinfo {year} {2016})}\BibitemShut {NoStop}%
\bibitem [{\citenamefont {Guillam\'{o}n}\ \emph {et~al.}(2008)\citenamefont
  {Guillam\'{o}n}, \citenamefont {Suderow}, \citenamefont {Vieira},
  \citenamefont {Cario}, \citenamefont {Diener},\ and\ \citenamefont
  {Rodi\`{e}re}}]{guillamon_superconducting_2008}%
  \BibitemOpen
  \bibfield  {author} {\bibinfo {author} {\bibfnamefont {I.}~\bibnamefont
  {Guillam\'{o}n}}, \bibinfo {author} {\bibfnamefont {H.}~\bibnamefont
  {Suderow}}, \bibinfo {author} {\bibfnamefont {S.}~\bibnamefont {Vieira}},
  \bibinfo {author} {\bibfnamefont {L.}~\bibnamefont {Cario}}, \bibinfo
  {author} {\bibfnamefont {P.}~\bibnamefont {Diener}}, \ and\ \bibinfo {author}
  {\bibfnamefont {P.}~\bibnamefont {Rodi\`{e}re}},\ }\bibfield  {title}
  {\enquote {\bibinfo {title} {Superconducting {Density} of {States} and
  {Vortex} {Cores} of {2H-NbS}$_2$},}\ }\href {\doibase
  10.1103/PhysRevLett.101.166407} {\bibfield  {journal} {\bibinfo  {journal}
  {Phys. Rev. Lett.}\ }\textbf {\bibinfo {volume} {101}},\ \bibinfo {pages}
  {166407} (\bibinfo {year} {2008})}\BibitemShut {NoStop}%
\bibitem [{\citenamefont {Tissen}\ \emph {et~al.}(2013)\citenamefont {Tissen},
  \citenamefont {Osorio}, \citenamefont {Brison}, \citenamefont {Nemes},
  \citenamefont {Garc\'{\i}a-Hern\'andez}, \citenamefont {Cario}, \citenamefont
  {Rodi\`ere}, \citenamefont {Vieira},\ and\ \citenamefont
  {Suderow}}]{tissen_pressure_2013}%
  \BibitemOpen
  \bibfield  {author} {\bibinfo {author} {\bibfnamefont {V.~G.}\ \bibnamefont
  {Tissen}}, \bibinfo {author} {\bibfnamefont {M.~R.}\ \bibnamefont {Osorio}},
  \bibinfo {author} {\bibfnamefont {J.~P.}\ \bibnamefont {Brison}}, \bibinfo
  {author} {\bibfnamefont {N.~M.}\ \bibnamefont {Nemes}}, \bibinfo {author}
  {\bibfnamefont {M.}~\bibnamefont {Garc\'{\i}a-Hern\'andez}}, \bibinfo
  {author} {\bibfnamefont {L.}~\bibnamefont {Cario}}, \bibinfo {author}
  {\bibfnamefont {P.}~\bibnamefont {Rodi\`ere}}, \bibinfo {author}
  {\bibfnamefont {S.}~\bibnamefont {Vieira}}, \ and\ \bibinfo {author}
  {\bibfnamefont {H.}~\bibnamefont {Suderow}},\ }\bibfield  {title} {\enquote
  {\bibinfo {title} {Pressure dependence of superconducting critical
  temperature and upper critical field of {2H-NbS}${}_{2}$},}\ }\href {\doibase
  10.1103/PhysRevB.87.134502} {\bibfield  {journal} {\bibinfo  {journal} {Phys.
  Rev. B}\ }\textbf {\bibinfo {volume} {87}},\ \bibinfo {pages} {134502}
  (\bibinfo {year} {2013})}\BibitemShut {NoStop}%
\bibitem [{\citenamefont {Heil}\ \emph {et~al.}(2017)\citenamefont {Heil},
  \citenamefont {Ponc\'e}, \citenamefont {Lambert}, \citenamefont {Schlipf},
  \citenamefont {Margine},\ and\ \citenamefont {Giustino}}]{Heil17}%
  \BibitemOpen
  \bibfield  {author} {\bibinfo {author} {\bibfnamefont {Christoph}\
  \bibnamefont {Heil}}, \bibinfo {author} {\bibfnamefont {Samuel}\ \bibnamefont
  {Ponc\'e}}, \bibinfo {author} {\bibfnamefont {Henry}\ \bibnamefont
  {Lambert}}, \bibinfo {author} {\bibfnamefont {Martin}\ \bibnamefont
  {Schlipf}}, \bibinfo {author} {\bibfnamefont {Elena~R.}\ \bibnamefont
  {Margine}}, \ and\ \bibinfo {author} {\bibfnamefont {Feliciano}\ \bibnamefont
  {Giustino}},\ }\bibfield  {title} {\enquote {\bibinfo {title} {Origin of
  superconductivity and latent charge density wave in {NbS}$_{2}$},}\ }\href
  {\doibase 10.1103/PhysRevLett.119.087003} {\bibfield  {journal} {\bibinfo
  {journal} {Phys. Rev. Lett.}\ }\textbf {\bibinfo {volume} {119}},\ \bibinfo
  {pages} {087003} (\bibinfo {year} {2017})}\BibitemShut {NoStop}%
\bibitem [{\citenamefont {Zhao}\ \emph {et~al.}(2016)\citenamefont {Zhao},
  \citenamefont {Hotta}, \citenamefont {Koretsune}, \citenamefont {Watanabe},
  \citenamefont {Taniguchi}, \citenamefont {Sugawara}, \citenamefont {{Takashi
  Takahashi}}, \citenamefont {Shinohara},\ and\ \citenamefont
  {Kitaura}}]{zhao_two-dimensional_2016}%
  \BibitemOpen
  \bibfield  {author} {\bibinfo {author} {\bibfnamefont {Sihan}\ \bibnamefont
  {Zhao}}, \bibinfo {author} {\bibfnamefont {Takato}\ \bibnamefont {Hotta}},
  \bibinfo {author} {\bibfnamefont {Takashi}\ \bibnamefont {Koretsune}},
  \bibinfo {author} {\bibfnamefont {Kenji}\ \bibnamefont {Watanabe}}, \bibinfo
  {author} {\bibfnamefont {Takashi}\ \bibnamefont {Taniguchi}}, \bibinfo
  {author} {\bibfnamefont {Katsuaki}\ \bibnamefont {Sugawara}}, \bibinfo
  {author} {\bibnamefont {{Takashi Takahashi}}}, \bibinfo {author}
  {\bibfnamefont {Hisanori}\ \bibnamefont {Shinohara}}, \ and\ \bibinfo
  {author} {\bibfnamefont {Ryo}\ \bibnamefont {Kitaura}},\ }\bibfield  {title}
  {\enquote {\bibinfo {title} {Two-dimensional metallic {NbS}$_2$ : growth,
  optical identification and transport properties},}\ }\href {\doibase
  10.1088/2053-1583/3/2/025027} {\bibfield  {journal} {\bibinfo  {journal} {2D
  Mater.}\ }\textbf {\bibinfo {volume} {3}},\ \bibinfo {pages} {025027}
  (\bibinfo {year} {2016})}\BibitemShut {NoStop}%
\bibitem [{\citenamefont {Xu}\ \emph {et~al.}(2014)\citenamefont {Xu},
  \citenamefont {Liu},\ and\ \citenamefont {Guo}}]{xu_tensile_2014}%
  \BibitemOpen
  \bibfield  {author} {\bibinfo {author} {\bibfnamefont {Ying}\ \bibnamefont
  {Xu}}, \bibinfo {author} {\bibfnamefont {Xiaofei}\ \bibnamefont {Liu}}, \
  and\ \bibinfo {author} {\bibfnamefont {Wanlin}\ \bibnamefont {Guo}},\
  }\bibfield  {title} {\enquote {\bibinfo {title} {Tensile strain induced
  switching of magnetic states in {NbSe}$_2$ and {NbS}$_2$ single layers},}\
  }\href {\doibase 10.1039/C4NR01486C} {\bibfield  {journal} {\bibinfo
  {journal} {Nanoscale}\ }\textbf {\bibinfo {volume} {6}},\ \bibinfo {pages}
  {12929--12933} (\bibinfo {year} {2014})}\BibitemShut {NoStop}%
\bibitem [{\citenamefont {G{\"u}ller}\ \emph {et~al.}(2016)\citenamefont
  {G{\"u}ller}, \citenamefont {Vildosola},\ and\ \citenamefont
  {Llois}}]{guller_spin_2016}%
  \BibitemOpen
  \bibfield  {author} {\bibinfo {author} {\bibfnamefont {F.}~\bibnamefont
  {G{\"u}ller}}, \bibinfo {author} {\bibfnamefont {V.~L.}\ \bibnamefont
  {Vildosola}}, \ and\ \bibinfo {author} {\bibfnamefont {A.~M.}\ \bibnamefont
  {Llois}},\ }\bibfield  {title} {\enquote {\bibinfo {title} {Spin density wave
  instabilities in the {NbS}$_2$ monolayer},}\ }\href {\doibase
  10.1103/PhysRevB.93.094434} {\bibfield  {journal} {\bibinfo  {journal} {Phys.
  Rev. B}\ }\textbf {\bibinfo {volume} {93}},\ \bibinfo {pages} {094434}
  (\bibinfo {year} {2016})}\BibitemShut {NoStop}%
\bibitem [{\citenamefont {Rubtsov}\ \emph {et~al.}(2012)\citenamefont
  {Rubtsov}, \citenamefont {Katsnelson},\ and\ \citenamefont
  {Lichtenstein}}]{Rubtsov12}%
  \BibitemOpen
  \bibfield  {author} {\bibinfo {author} {\bibfnamefont {A.~N.}\ \bibnamefont
  {Rubtsov}}, \bibinfo {author} {\bibfnamefont {M.~I.}\ \bibnamefont
  {Katsnelson}}, \ and\ \bibinfo {author} {\bibfnamefont {A.~I.}\ \bibnamefont
  {Lichtenstein}},\ }\bibfield  {title} {\enquote {\bibinfo {title} {Dual boson
  approach to collective excitations in correlated fermionic systems},}\ }\href
  {\doibase 10.1016/j.aop.2012.01.002} {\bibfield  {journal} {\bibinfo
  {journal} {Annals of Physics}\ }\textbf {\bibinfo {volume} {327}},\ \bibinfo
  {pages} {1320} (\bibinfo {year} {2012})}\BibitemShut {NoStop}%
\bibitem [{\citenamefont {van Loon}\ \emph
  {et~al.}(2014{\natexlab{a}})\citenamefont {van Loon}, \citenamefont
  {Hafermann}, \citenamefont {Lichtenstein}, \citenamefont {Rubtsov},\ and\
  \citenamefont {Katsnelson}}]{vanLoon14}%
  \BibitemOpen
  \bibfield  {author} {\bibinfo {author} {\bibfnamefont {E.~G. C.~P.}\
  \bibnamefont {van Loon}}, \bibinfo {author} {\bibfnamefont {H.}~\bibnamefont
  {Hafermann}}, \bibinfo {author} {\bibfnamefont {A.~I.}\ \bibnamefont
  {Lichtenstein}}, \bibinfo {author} {\bibfnamefont {A.~N.}\ \bibnamefont
  {Rubtsov}}, \ and\ \bibinfo {author} {\bibfnamefont {M.~I.}\ \bibnamefont
  {Katsnelson}},\ }\bibfield  {title} {\enquote {\bibinfo {title} {Plasmons in
  strongly correlated systems: Spectral weight transfer and renormalized
  dispersion},}\ }\href {\doibase 10.1103/PhysRevLett.113.246407} {\bibfield
  {journal} {\bibinfo  {journal} {Phys. Rev. Lett.}\ }\textbf {\bibinfo
  {volume} {113}},\ \bibinfo {pages} {246407} (\bibinfo {year}
  {2014}{\natexlab{a}})}\BibitemShut {NoStop}%
\bibitem [{\citenamefont {Hafermann}\ \emph {et~al.}(2014)\citenamefont
  {Hafermann}, \citenamefont {van Loon}, \citenamefont {Katsnelson},
  \citenamefont {Lichtenstein},\ and\ \citenamefont
  {Parcollet}}]{Hafermann14-2}%
  \BibitemOpen
  \bibfield  {author} {\bibinfo {author} {\bibfnamefont {Hartmut}\ \bibnamefont
  {Hafermann}}, \bibinfo {author} {\bibfnamefont {Erik G. C.~P.}\ \bibnamefont
  {van Loon}}, \bibinfo {author} {\bibfnamefont {Mikhail~I.}\ \bibnamefont
  {Katsnelson}}, \bibinfo {author} {\bibfnamefont {Alexander~I.}\ \bibnamefont
  {Lichtenstein}}, \ and\ \bibinfo {author} {\bibfnamefont {Olivier}\
  \bibnamefont {Parcollet}},\ }\bibfield  {title} {\enquote {\bibinfo {title}
  {Collective charge excitations of strongly correlated electrons, vertex
  corrections, and gauge invariance},}\ }\href {\doibase
  10.1103/PhysRevB.90.235105} {\bibfield  {journal} {\bibinfo  {journal} {Phys.
  Rev. B}\ }\textbf {\bibinfo {volume} {90}},\ \bibinfo {pages} {235105}
  (\bibinfo {year} {2014})}\BibitemShut {NoStop}%
\bibitem [{\citenamefont {Berger}\ \emph {et~al.}(1995)\citenamefont {Berger},
  \citenamefont {Valasek},\ and\ \citenamefont {von~der
  Linden}}]{berger_two-dimensional_1995}%
  \BibitemOpen
  \bibfield  {author} {\bibinfo {author} {\bibfnamefont {E.}~\bibnamefont
  {Berger}}, \bibinfo {author} {\bibfnamefont {P.}~\bibnamefont {Valasek}}, \
  and\ \bibinfo {author} {\bibfnamefont {W.}~\bibnamefont {von~der Linden}},\
  }\bibfield  {title} {\enquote {\bibinfo {title} {Two-dimensional
  {Hubbard}-{Holstein} model},}\ }\href {\doibase 10.1103/PhysRevB.52.4806}
  {\bibfield  {journal} {\bibinfo  {journal} {Phys. Rev. B}\ }\textbf {\bibinfo
  {volume} {52}},\ \bibinfo {pages} {4806--4814} (\bibinfo {year}
  {1995})}\BibitemShut {NoStop}%
\bibitem [{\citenamefont {Sangiovanni}\ \emph {et~al.}(2005)\citenamefont
  {Sangiovanni}, \citenamefont {Capone}, \citenamefont {Castellani},\ and\
  \citenamefont {Grilli}}]{PhysRevLett.94.026401}%
  \BibitemOpen
  \bibfield  {author} {\bibinfo {author} {\bibfnamefont {G.}~\bibnamefont
  {Sangiovanni}}, \bibinfo {author} {\bibfnamefont {M.}~\bibnamefont {Capone}},
  \bibinfo {author} {\bibfnamefont {C.}~\bibnamefont {Castellani}}, \ and\
  \bibinfo {author} {\bibfnamefont {M.}~\bibnamefont {Grilli}},\ }\bibfield
  {title} {\enquote {\bibinfo {title} {Electron-phonon interaction close to a
  {Mott} transition},}\ }\href {\doibase 10.1103/PhysRevLett.94.026401}
  {\bibfield  {journal} {\bibinfo  {journal} {Phys. Rev. Lett.}\ }\textbf
  {\bibinfo {volume} {94}},\ \bibinfo {pages} {026401} (\bibinfo {year}
  {2005})}\BibitemShut {NoStop}%
\bibitem [{\citenamefont {Werner}\ and\ \citenamefont
  {Millis}(2007)}]{PhysRevLett.99.146404}%
  \BibitemOpen
  \bibfield  {author} {\bibinfo {author} {\bibfnamefont {Philipp}\ \bibnamefont
  {Werner}}\ and\ \bibinfo {author} {\bibfnamefont {Andrew~J.}\ \bibnamefont
  {Millis}},\ }\bibfield  {title} {\enquote {\bibinfo {title} {Efficient
  dynamical mean field simulation of the {Holstein-Hubbard} model},}\ }\href
  {\doibase 10.1103/PhysRevLett.99.146404} {\bibfield  {journal} {\bibinfo
  {journal} {Phys. Rev. Lett.}\ }\textbf {\bibinfo {volume} {99}},\ \bibinfo
  {pages} {146404} (\bibinfo {year} {2007})}\BibitemShut {NoStop}%
\bibitem [{\citenamefont {Aryasetiawan}\ \emph {et~al.}(2004)\citenamefont
  {Aryasetiawan}, \citenamefont {Imada}, \citenamefont {Georges}, \citenamefont
  {Kotliar}, \citenamefont {Biermann},\ and\ \citenamefont
  {Lichtenstein}}]{PhysRevB.70.195104}%
  \BibitemOpen
  \bibfield  {author} {\bibinfo {author} {\bibfnamefont {F.}~\bibnamefont
  {Aryasetiawan}}, \bibinfo {author} {\bibfnamefont {M.}~\bibnamefont {Imada}},
  \bibinfo {author} {\bibfnamefont {A.}~\bibnamefont {Georges}}, \bibinfo
  {author} {\bibfnamefont {G.}~\bibnamefont {Kotliar}}, \bibinfo {author}
  {\bibfnamefont {S.}~\bibnamefont {Biermann}}, \ and\ \bibinfo {author}
  {\bibfnamefont {A.~I.}\ \bibnamefont {Lichtenstein}},\ }\bibfield  {title}
  {\enquote {\bibinfo {title} {Frequency-dependent local interactions and
  low-energy effective models from electronic structure calculations},}\ }\href
  {\doibase 10.1103/PhysRevB.70.195104} {\bibfield  {journal} {\bibinfo
  {journal} {Phys. Rev. B}\ }\textbf {\bibinfo {volume} {70}},\ \bibinfo
  {pages} {195104} (\bibinfo {year} {2004})}\BibitemShut {NoStop}%
\bibitem [{\citenamefont {Sch\"uler}\ \emph {et~al.}(2013)\citenamefont
  {Sch\"uler}, \citenamefont {R\"osner}, \citenamefont {Wehling}, \citenamefont
  {Lichtenstein},\ and\ \citenamefont {Katsnelson}}]{PhysRevLett.111.036601}%
  \BibitemOpen
  \bibfield  {author} {\bibinfo {author} {\bibfnamefont {M.}~\bibnamefont
  {Sch\"uler}}, \bibinfo {author} {\bibfnamefont {M.}~\bibnamefont {R\"osner}},
  \bibinfo {author} {\bibfnamefont {T.~O.}\ \bibnamefont {Wehling}}, \bibinfo
  {author} {\bibfnamefont {A.~I.}\ \bibnamefont {Lichtenstein}}, \ and\
  \bibinfo {author} {\bibfnamefont {M.~I.}\ \bibnamefont {Katsnelson}},\
  }\bibfield  {title} {\enquote {\bibinfo {title} {Optimal {Hubbard} models for
  materials with nonlocal {Coulomb} interactions: Graphene, silicene, and
  benzene},}\ }\href {\doibase 10.1103/PhysRevLett.111.036601} {\bibfield
  {journal} {\bibinfo  {journal} {Phys. Rev. Lett.}\ }\textbf {\bibinfo
  {volume} {111}},\ \bibinfo {pages} {036601} (\bibinfo {year}
  {2013})}\BibitemShut {NoStop}%
\bibitem [{\citenamefont {Sanders}\ \emph {et~al.}(2016)\citenamefont
  {Sanders}, \citenamefont {Dendzik}, \citenamefont {Ngankeu}, \citenamefont
  {Eich}, \citenamefont {Bruix}, \citenamefont {Bianchi}, \citenamefont {Miwa},
  \citenamefont {Hammer}, \citenamefont {Khajetoorians},\ and\ \citenamefont
  {Hofmann}}]{Sanders16}%
  \BibitemOpen
  \bibfield  {author} {\bibinfo {author} {\bibfnamefont {Charlotte~E.}\
  \bibnamefont {Sanders}}, \bibinfo {author} {\bibfnamefont {Maciej}\
  \bibnamefont {Dendzik}}, \bibinfo {author} {\bibfnamefont {Arlette~S.}\
  \bibnamefont {Ngankeu}}, \bibinfo {author} {\bibfnamefont {Andreas}\
  \bibnamefont {Eich}}, \bibinfo {author} {\bibfnamefont {Albert}\ \bibnamefont
  {Bruix}}, \bibinfo {author} {\bibfnamefont {Marco}\ \bibnamefont {Bianchi}},
  \bibinfo {author} {\bibfnamefont {Jill~A.}\ \bibnamefont {Miwa}}, \bibinfo
  {author} {\bibfnamefont {Bj\o{}rk}\ \bibnamefont {Hammer}}, \bibinfo {author}
  {\bibfnamefont {Alexander~A.}\ \bibnamefont {Khajetoorians}}, \ and\ \bibinfo
  {author} {\bibfnamefont {Philip}\ \bibnamefont {Hofmann}},\ }\bibfield
  {title} {\enquote {\bibinfo {title} {Crystalline and electronic structure of
  single-layer {TaS}$_{2}$},}\ }\href {\doibase 10.1103/PhysRevB.94.081404}
  {\bibfield  {journal} {\bibinfo  {journal} {Phys. Rev. B}\ }\textbf {\bibinfo
  {volume} {94}},\ \bibinfo {pages} {081404} (\bibinfo {year}
  {2016})}\BibitemShut {NoStop}%
\bibitem [{\citenamefont {Jeon}\ \emph {et~al.}(2004)\citenamefont {Jeon},
  \citenamefont {Park}, \citenamefont {Han}, \citenamefont {Lee},\ and\
  \citenamefont {Choi}}]{Jeon04}%
  \BibitemOpen
  \bibfield  {author} {\bibinfo {author} {\bibfnamefont {Gun~Sang}\
  \bibnamefont {Jeon}}, \bibinfo {author} {\bibfnamefont {Tae-Ho}\ \bibnamefont
  {Park}}, \bibinfo {author} {\bibfnamefont {Jung~Hoon}\ \bibnamefont {Han}},
  \bibinfo {author} {\bibfnamefont {Hyun~C.}\ \bibnamefont {Lee}}, \ and\
  \bibinfo {author} {\bibfnamefont {Han-Yong}\ \bibnamefont {Choi}},\
  }\bibfield  {title} {\enquote {\bibinfo {title} {Dynamical mean-field theory
  of the {Hubbard-Holstein} model at half filling: Zero temperature
  metal-insulator and insulator-insulator transitions},}\ }\href {\doibase
  10.1103/PhysRevB.70.125114} {\bibfield  {journal} {\bibinfo  {journal} {Phys.
  Rev. B}\ }\textbf {\bibinfo {volume} {70}},\ \bibinfo {pages} {125114}
  (\bibinfo {year} {2004})}\BibitemShut {NoStop}%
\bibitem [{\citenamefont {Yoshioka}\ \emph {et~al.}(2010)\citenamefont
  {Yoshioka}, \citenamefont {Koga},\ and\ \citenamefont
  {Kawakami}}]{Yoshioka10}%
  \BibitemOpen
  \bibfield  {author} {\bibinfo {author} {\bibfnamefont {Takuya}\ \bibnamefont
  {Yoshioka}}, \bibinfo {author} {\bibfnamefont {Akihisa}\ \bibnamefont
  {Koga}}, \ and\ \bibinfo {author} {\bibfnamefont {Norio}\ \bibnamefont
  {Kawakami}},\ }\bibfield  {title} {\enquote {\bibinfo {title} {Mott
  transition in the {H}ubbard model on the triangular lattice},}\ }\href
  {\doibase 10.1002/pssb.200983020} {\bibfield  {journal} {\bibinfo  {journal}
  {Physica Status Solidi (b)}\ }\textbf {\bibinfo {volume} {247}},\ \bibinfo
  {pages} {635--637} (\bibinfo {year} {2010})}\BibitemShut {NoStop}%
\bibitem [{\citenamefont {{Tranquada J. M.}}\ \emph {et~al.}(1995)\citenamefont
  {{Tranquada J. M.}}, \citenamefont {{Sternlieb B. J.}}, \citenamefont {{Axe
  J. D.}}, \citenamefont {{Nakamura Y.}},\ and\ \citenamefont {{Uchida
  S.}}}]{Tranquada95}%
  \BibitemOpen
  \bibfield  {author} {\bibinfo {author} {\bibnamefont {{Tranquada J. M.}}},
  \bibinfo {author} {\bibnamefont {{Sternlieb B. J.}}}, \bibinfo {author}
  {\bibnamefont {{Axe J. D.}}}, \bibinfo {author} {\bibnamefont {{Nakamura
  Y.}}}, \ and\ \bibinfo {author} {\bibnamefont {{Uchida S.}}},\ }\bibfield
  {title} {\enquote {\bibinfo {title} {{Evidence for stripe correlations of
  spins and holes in copper oxide superconductors}},}\ }\href {\doibase
  http://dx.doi.org/10.1038/375561a0} {\bibfield  {journal} {\bibinfo
  {journal} {Nature}\ }\textbf {\bibinfo {volume} {375}},\ \bibinfo {pages}
  {561--563} (\bibinfo {year} {1995})},\ \bibinfo {note}
  {10.1038/375561a0}\BibitemShut {NoStop}%
\bibitem [{\citenamefont {Hansmann}\ \emph {et~al.}(2013)\citenamefont
  {Hansmann}, \citenamefont {Ayral}, \citenamefont {Vaugier}, \citenamefont
  {Werner},\ and\ \citenamefont {Biermann}}]{Hansmann13}%
  \BibitemOpen
  \bibfield  {author} {\bibinfo {author} {\bibfnamefont {P.}~\bibnamefont
  {Hansmann}}, \bibinfo {author} {\bibfnamefont {T.}~\bibnamefont {Ayral}},
  \bibinfo {author} {\bibfnamefont {L.}~\bibnamefont {Vaugier}}, \bibinfo
  {author} {\bibfnamefont {P.}~\bibnamefont {Werner}}, \ and\ \bibinfo {author}
  {\bibfnamefont {S.}~\bibnamefont {Biermann}},\ }\bibfield  {title} {\enquote
  {\bibinfo {title} {Long-range coulomb interactions in surface systems: A
  first-principles description within self-consistently combined {$GW$} and
  dynamical mean-field theory},}\ }\href {\doibase
  10.1103/PhysRevLett.110.166401} {\bibfield  {journal} {\bibinfo  {journal}
  {Phys. Rev. Lett.}\ }\textbf {\bibinfo {volume} {110}},\ \bibinfo {pages}
  {166401} (\bibinfo {year} {2013})}\BibitemShut {NoStop}%
\bibitem [{\citenamefont {Werner}\ \emph {et~al.}(2008)\citenamefont {Werner},
  \citenamefont {Gull}, \citenamefont {Troyer},\ and\ \citenamefont
  {Millis}}]{Werner08}%
  \BibitemOpen
  \bibfield  {author} {\bibinfo {author} {\bibfnamefont {Philipp}\ \bibnamefont
  {Werner}}, \bibinfo {author} {\bibfnamefont {Emanuel}\ \bibnamefont {Gull}},
  \bibinfo {author} {\bibfnamefont {Matthias}\ \bibnamefont {Troyer}}, \ and\
  \bibinfo {author} {\bibfnamefont {Andrew~J.}\ \bibnamefont {Millis}},\
  }\bibfield  {title} {\enquote {\bibinfo {title} {Spin freezing transition and
  non-fermi-liquid self-energy in a three-orbital model},}\ }\href {\doibase
  10.1103/PhysRevLett.101.166405} {\bibfield  {journal} {\bibinfo  {journal}
  {Phys. Rev. Lett.}\ }\textbf {\bibinfo {volume} {101}},\ \bibinfo {pages}
  {166405} (\bibinfo {year} {2008})}\BibitemShut {NoStop}%
\bibitem [{\citenamefont {Haule}\ and\ \citenamefont
  {Kotliar}(2009)}]{Haule09}%
  \BibitemOpen
  \bibfield  {author} {\bibinfo {author} {\bibfnamefont {K}~\bibnamefont
  {Haule}}\ and\ \bibinfo {author} {\bibfnamefont {G}~\bibnamefont {Kotliar}},\
  }\bibfield  {title} {\enquote {\bibinfo {title} {Coherence–incoherence
  crossover in the normal state of iron oxypnictides and importance of {Hund's}
  rule coupling},}\ }\href {http://stacks.iop.org/1367-2630/11/i=2/a=025021}
  {\bibfield  {journal} {\bibinfo  {journal} {New Journal of Physics}\ }\textbf
  {\bibinfo {volume} {11}},\ \bibinfo {pages} {025021} (\bibinfo {year}
  {2009})}\BibitemShut {NoStop}%
\bibitem [{\citenamefont {{Yin Z. P.}}\ \emph {et~al.}(2011)\citenamefont {{Yin
  Z. P.}}, \citenamefont {{Haule K.}},\ and\ \citenamefont {{Kotliar
  G.}}}]{Yin11}%
  \BibitemOpen
  \bibfield  {author} {\bibinfo {author} {\bibnamefont {{Yin Z. P.}}}, \bibinfo
  {author} {\bibnamefont {{Haule K.}}}, \ and\ \bibinfo {author} {\bibnamefont
  {{Kotliar G.}}},\ }\bibfield  {title} {\enquote {\bibinfo {title} {{Kinetic
  frustration and the nature of the magnetic and paramagnetic states in iron
  pnictides and iron chalcogenides}},}\ }\href {\doibase
  http://dx.doi.org/10.1038/nmat3120 10.1038/nmat3120} {\bibfield  {journal}
  {\bibinfo  {journal} {Nature Materials}\ }\textbf {\bibinfo {volume} {10}},\
  \bibinfo {pages} {932} (\bibinfo {year} {2011})}\BibitemShut {NoStop}%
\bibitem [{\citenamefont {de' Medici}\ \emph {et~al.}(2011)\citenamefont {de'
  Medici}, \citenamefont {Mravlje},\ and\ \citenamefont {Georges}}]{Medici11}%
  \BibitemOpen
  \bibfield  {author} {\bibinfo {author} {\bibfnamefont {Luca}\ \bibnamefont
  {de' Medici}}, \bibinfo {author} {\bibfnamefont {Jernej}\ \bibnamefont
  {Mravlje}}, \ and\ \bibinfo {author} {\bibfnamefont {Antoine}\ \bibnamefont
  {Georges}},\ }\bibfield  {title} {\enquote {\bibinfo {title} {Janus-faced
  influence of {Hund's} rule coupling in strongly correlated materials},}\
  }\href {\doibase 10.1103/PhysRevLett.107.256401} {\bibfield  {journal}
  {\bibinfo  {journal} {Phys. Rev. Lett.}\ }\textbf {\bibinfo {volume} {107}},\
  \bibinfo {pages} {256401} (\bibinfo {year} {2011})}\BibitemShut {NoStop}%
\bibitem [{\citenamefont {Georges}\ \emph {et~al.}(2013)\citenamefont
  {Georges}, \citenamefont {de' Medici},\ and\ \citenamefont
  {Mravlje}}]{Georges13}%
  \BibitemOpen
  \bibfield  {author} {\bibinfo {author} {\bibfnamefont {Antoine}\ \bibnamefont
  {Georges}}, \bibinfo {author} {\bibfnamefont {Luca}\ \bibnamefont {de'
  Medici}}, \ and\ \bibinfo {author} {\bibfnamefont {Jernej}\ \bibnamefont
  {Mravlje}},\ }\bibfield  {title} {\enquote {\bibinfo {title} {Strong
  correlations from {Hund's} coupling},}\ }\href {\doibase
  10.1146/annurev-conmatphys-020911-125045} {\bibfield  {journal} {\bibinfo
  {journal} {Annual Review of Condensed Matter Physics}\ }\textbf {\bibinfo
  {volume} {4}},\ \bibinfo {pages} {137--178} (\bibinfo {year}
  {2013})}\BibitemShut {NoStop}%
\bibitem [{\citenamefont {Ament}\ \emph {et~al.}(2011)\citenamefont {Ament},
  \citenamefont {van Veenendaal}, \citenamefont {Devereaux}, \citenamefont
  {Hill},\ and\ \citenamefont {van~den Brink}}]{Ament11}%
  \BibitemOpen
  \bibfield  {author} {\bibinfo {author} {\bibfnamefont {Luuk J.~P.}\
  \bibnamefont {Ament}}, \bibinfo {author} {\bibfnamefont {Michel}\
  \bibnamefont {van Veenendaal}}, \bibinfo {author} {\bibfnamefont {Thomas~P.}\
  \bibnamefont {Devereaux}}, \bibinfo {author} {\bibfnamefont {John~P.}\
  \bibnamefont {Hill}}, \ and\ \bibinfo {author} {\bibfnamefont {Jeroen}\
  \bibnamefont {van~den Brink}},\ }\bibfield  {title} {\enquote {\bibinfo
  {title} {Resonant inelastic x-ray scattering studies of elementary
  excitations},}\ }\href {\doibase 10.1103/RevModPhys.83.705} {\bibfield
  {journal} {\bibinfo  {journal} {Rev. Mod. Phys.}\ }\textbf {\bibinfo {volume}
  {83}},\ \bibinfo {pages} {705--767} (\bibinfo {year} {2011})}\BibitemShut
  {NoStop}%
\bibitem [{\citenamefont {Vig}\ \emph {et~al.}(2017)\citenamefont {Vig},
  \citenamefont {Kogar}, \citenamefont {Mitrano}, \citenamefont {Husain},
  \citenamefont {Mishra}, \citenamefont {Rak}, \citenamefont {Venema},
  \citenamefont {Johnson}, \citenamefont {Gu}, \citenamefont {Fradkin},
  \citenamefont {Norman},\ and\ \citenamefont {Abbamonte}}]{Vig17}%
  \BibitemOpen
  \bibfield  {author} {\bibinfo {author} {\bibfnamefont {Sean}\ \bibnamefont
  {Vig}}, \bibinfo {author} {\bibfnamefont {Anshul}\ \bibnamefont {Kogar}},
  \bibinfo {author} {\bibfnamefont {Matteo}\ \bibnamefont {Mitrano}}, \bibinfo
  {author} {\bibfnamefont {Ali~A.}\ \bibnamefont {Husain}}, \bibinfo {author}
  {\bibfnamefont {Vivek}\ \bibnamefont {Mishra}}, \bibinfo {author}
  {\bibfnamefont {Melinda~S.}\ \bibnamefont {Rak}}, \bibinfo {author}
  {\bibfnamefont {Luc}\ \bibnamefont {Venema}}, \bibinfo {author}
  {\bibfnamefont {Peter~D.}\ \bibnamefont {Johnson}}, \bibinfo {author}
  {\bibfnamefont {Genda~D.}\ \bibnamefont {Gu}}, \bibinfo {author}
  {\bibfnamefont {Eduardo}\ \bibnamefont {Fradkin}}, \bibinfo {author}
  {\bibfnamefont {Michael~R.}\ \bibnamefont {Norman}}, \ and\ \bibinfo {author}
  {\bibfnamefont {Peter}\ \bibnamefont {Abbamonte}},\ }\bibfield  {title}
  {\enquote {\bibinfo {title} {{Measurement of the dynamic charge response of
  materials using low-energy, momentum-resolved electron energy-loss
  spectroscopy (M-EELS)}},}\ }\href {\doibase 10.21468/SciPostPhys.3.4.026}
  {\bibfield  {journal} {\bibinfo  {journal} {SciPost Phys.}\ }\textbf
  {\bibinfo {volume} {3}},\ \bibinfo {pages} {026} (\bibinfo {year}
  {2017})}\BibitemShut {NoStop}%
\bibitem [{\citenamefont {{Menard Gerbold C.}}\ \emph
  {et~al.}(2015)\citenamefont {{Menard Gerbold C.}}, \citenamefont {{Guissart
  Sebastien}}, \citenamefont {{Brun Christophe}}, \citenamefont {{Pons
  Stephane}}, \citenamefont {{Stolyarov Vasily S.}}, \citenamefont
  {{Debontridder Francois}}, \citenamefont {{Leclerc Matthieu V.}},
  \citenamefont {{Janod Etienne}}, \citenamefont {{Cario Laurent}},
  \citenamefont {{Roditchev Dimitri}}, \citenamefont {{Simon Pascal}},\ and\
  \citenamefont {{Cren Tristan}}}]{Menard15}%
  \BibitemOpen
  \bibfield  {author} {\bibinfo {author} {\bibnamefont {{Menard Gerbold C.}}},
  \bibinfo {author} {\bibnamefont {{Guissart Sebastien}}}, \bibinfo {author}
  {\bibnamefont {{Brun Christophe}}}, \bibinfo {author} {\bibnamefont {{Pons
  Stephane}}}, \bibinfo {author} {\bibnamefont {{Stolyarov Vasily S.}}},
  \bibinfo {author} {\bibnamefont {{Debontridder Francois}}}, \bibinfo {author}
  {\bibnamefont {{Leclerc Matthieu V.}}}, \bibinfo {author} {\bibnamefont
  {{Janod Etienne}}}, \bibinfo {author} {\bibnamefont {{Cario Laurent}}},
  \bibinfo {author} {\bibnamefont {{Roditchev Dimitri}}}, \bibinfo {author}
  {\bibnamefont {{Simon Pascal}}}, \ and\ \bibinfo {author} {\bibnamefont
  {{Cren Tristan}}},\ }\bibfield  {title} {\enquote {\bibinfo {title}
  {{Coherent long-range magnetic bound states in a superconductor}},}\ }\href
  {\doibase http://dx.doi.org/10.1038/nphys3508 10.1038/nphys3508} {\bibfield
  {journal} {\bibinfo  {journal} {Nat Phys}\ }\textbf {\bibinfo {volume}
  {11}},\ \bibinfo {pages} {1013--1016} (\bibinfo {year} {2015})}\BibitemShut
  {NoStop}%
\bibitem [{\citenamefont {{Kezilebieke Shawulienu}}\ \emph
  {et~al.}(2018)\citenamefont {{Kezilebieke Shawulienu}}, \citenamefont
  {{Dvorak Marc}}, \citenamefont {{Ojanen Teemu}},\ and\ \citenamefont
  {{Liljeroth Peter}}}]{Kezilebieke18}%
  \BibitemOpen
  \bibfield  {author} {\bibinfo {author} {\bibnamefont {{Kezilebieke
  Shawulienu}}}, \bibinfo {author} {\bibnamefont {{Dvorak Marc}}}, \bibinfo
  {author} {\bibnamefont {{Ojanen Teemu}}}, \ and\ \bibinfo {author}
  {\bibnamefont {{Liljeroth Peter}}},\ }\bibfield  {title} {\enquote {\bibinfo
  {title} {{Coupled Yu--Shiba--Rusinov States in Molecular Dimers on NbSe2}},}\
  }\href {\doibase https://doi.org/10.1021/acs.nanolett.7b05050} {\bibfield
  {journal} {\bibinfo  {journal} {Nano Letters}\ }\textbf {\bibinfo {volume}
  {18}},\ \bibinfo {pages} {2311--2315} (\bibinfo {year} {2018})},\ \bibinfo
  {note} {doi: 10.1021/acs.nanolett.7b05050}\BibitemShut {NoStop}%
\bibitem [{\citenamefont {Engel}\ and\ \citenamefont {van~den
  Broeck}(2001)}]{Engel2001}%
  \BibitemOpen
  \bibfield  {author} {\bibinfo {author} {\bibfnamefont {A}~\bibnamefont
  {Engel}}\ and\ \bibinfo {author} {\bibfnamefont {C}~\bibnamefont {van~den
  Broeck}},\ }\href@noop {} {\emph {\bibinfo {title} {Statistical mechanics of
  learning}}}\ (\bibinfo  {publisher} {Cambridge University Press,},\ \bibinfo
  {year} {2001})\BibitemShut {NoStop}%
\bibitem [{FLE(2014)}]{FLEUR}%
  \BibitemOpen
  \href@noop {} {\enquote {\bibinfo {title} {The {J\"ulich} {FLEUR} project},}\
  }\bibinfo {howpublished} {\url{http://www.flapw.de}} (\bibinfo {year}
  {2014})\BibitemShut {NoStop}%
\bibitem [{\citenamefont {Mostofi}\ \emph {et~al.}(2014)\citenamefont
  {Mostofi}, \citenamefont {Yates}, \citenamefont {Pizzi}, \citenamefont {Lee},
  \citenamefont {Souza}, \citenamefont {Vanderbilt},\ and\ \citenamefont
  {Marzari}}]{mostofi_updated_2014}%
  \BibitemOpen
  \bibfield  {author} {\bibinfo {author} {\bibfnamefont {Arash~A.}\
  \bibnamefont {Mostofi}}, \bibinfo {author} {\bibfnamefont {Jonathan~R.}\
  \bibnamefont {Yates}}, \bibinfo {author} {\bibfnamefont {Giovanni}\
  \bibnamefont {Pizzi}}, \bibinfo {author} {\bibfnamefont {Young-Su}\
  \bibnamefont {Lee}}, \bibinfo {author} {\bibfnamefont {Ivo}\ \bibnamefont
  {Souza}}, \bibinfo {author} {\bibfnamefont {David}\ \bibnamefont
  {Vanderbilt}}, \ and\ \bibinfo {author} {\bibfnamefont {Nicola}\ \bibnamefont
  {Marzari}},\ }\bibfield  {title} {\enquote {\bibinfo {title} {An updated
  version of wannier90: {A} tool for obtaining maximally-localised {Wannier}
  functions},}\ }\href {\doibase 10.1016/j.cpc.2014.05.003} {\bibfield
  {journal} {\bibinfo  {journal} {Comp. Phys. Comm.}\ }\textbf {\bibinfo
  {volume} {185}},\ \bibinfo {pages} {2309--2310} (\bibinfo {year}
  {2014})}\BibitemShut {NoStop}%
\bibitem [{\citenamefont {Baroni}\ \emph {et~al.}(2001)\citenamefont {Baroni},
  \citenamefont {de~Gironcoli}, \citenamefont {Dal~Corso},\ and\ \citenamefont
  {Giannozzi}}]{baroni_phonons_2001}%
  \BibitemOpen
  \bibfield  {author} {\bibinfo {author} {\bibfnamefont {Stefano}\ \bibnamefont
  {Baroni}}, \bibinfo {author} {\bibfnamefont {Stefano}\ \bibnamefont
  {de~Gironcoli}}, \bibinfo {author} {\bibfnamefont {Andrea}\ \bibnamefont
  {Dal~Corso}}, \ and\ \bibinfo {author} {\bibfnamefont {Paolo}\ \bibnamefont
  {Giannozzi}},\ }\bibfield  {title} {\enquote {\bibinfo {title} {Phonons and
  related crystal properties from density-functional perturbation theory},}\
  }\href {\doibase 10.1103/RevModPhys.73.515} {\bibfield  {journal} {\bibinfo
  {journal} {Rev. Mod. Phys.}\ }\textbf {\bibinfo {volume} {73}},\ \bibinfo
  {pages} {515--562} (\bibinfo {year} {2001})}\BibitemShut {NoStop}%
\bibitem [{\citenamefont {Giannozzi}\ \emph {et~al.}(2009)\citenamefont
  {Giannozzi}, \citenamefont {Baroni}, \citenamefont {Bonini}, \citenamefont
  {Calandra}, \citenamefont {Car}, \citenamefont {Cavazzoni}, \citenamefont
  {Ceresoli}, \citenamefont {Chiarotti}, \citenamefont {Cococcioni},
  \citenamefont {Dabo}, \citenamefont {Corso}, \citenamefont {Gironcoli},
  \citenamefont {Fabris}, \citenamefont {Fratesi}, \citenamefont {Gebauer},
  \citenamefont {Gerstmann}, \citenamefont {Gougoussis}, \citenamefont
  {Kokalj}, \citenamefont {Lazzeri}, \citenamefont {Martin-Samos},
  \citenamefont {Marzari}, \citenamefont {Mauri}, \citenamefont {Mazzarello},
  \citenamefont {Paolini}, \citenamefont {Pasquarello}, \citenamefont
  {Paulatto}, \citenamefont {Sbraccia}, \citenamefont {Scandolo}, \citenamefont
  {Sclauzero}, \citenamefont {Seitsonen}, \citenamefont {Smogunov},
  \citenamefont {Umari},\ and\ \citenamefont
  {Wentzcovitch}}]{giannozzi_quantum_2009}%
  \BibitemOpen
  \bibfield  {author} {\bibinfo {author} {\bibfnamefont {Paolo}\ \bibnamefont
  {Giannozzi}}, \bibinfo {author} {\bibfnamefont {Stefano}\ \bibnamefont
  {Baroni}}, \bibinfo {author} {\bibfnamefont {Nicola}\ \bibnamefont {Bonini}},
  \bibinfo {author} {\bibfnamefont {Matteo}\ \bibnamefont {Calandra}}, \bibinfo
  {author} {\bibfnamefont {Roberto}\ \bibnamefont {Car}}, \bibinfo {author}
  {\bibfnamefont {Carlo}\ \bibnamefont {Cavazzoni}}, \bibinfo {author}
  {\bibfnamefont {Davide}\ \bibnamefont {Ceresoli}}, \bibinfo {author}
  {\bibfnamefont {Guido~L.}\ \bibnamefont {Chiarotti}}, \bibinfo {author}
  {\bibfnamefont {Matteo}\ \bibnamefont {Cococcioni}}, \bibinfo {author}
  {\bibfnamefont {Ismaila}\ \bibnamefont {Dabo}}, \bibinfo {author}
  {\bibfnamefont {Andrea~Dal}\ \bibnamefont {Corso}}, \bibinfo {author}
  {\bibfnamefont {Stefano~de}\ \bibnamefont {Gironcoli}}, \bibinfo {author}
  {\bibfnamefont {Stefano}\ \bibnamefont {Fabris}}, \bibinfo {author}
  {\bibfnamefont {Guido}\ \bibnamefont {Fratesi}}, \bibinfo {author}
  {\bibfnamefont {Ralph}\ \bibnamefont {Gebauer}}, \bibinfo {author}
  {\bibfnamefont {Uwe}\ \bibnamefont {Gerstmann}}, \bibinfo {author}
  {\bibfnamefont {Christos}\ \bibnamefont {Gougoussis}}, \bibinfo {author}
  {\bibfnamefont {Anton}\ \bibnamefont {Kokalj}}, \bibinfo {author}
  {\bibfnamefont {Michele}\ \bibnamefont {Lazzeri}}, \bibinfo {author}
  {\bibfnamefont {Layla}\ \bibnamefont {Martin-Samos}}, \bibinfo {author}
  {\bibfnamefont {Nicola}\ \bibnamefont {Marzari}}, \bibinfo {author}
  {\bibfnamefont {Francesco}\ \bibnamefont {Mauri}}, \bibinfo {author}
  {\bibfnamefont {Riccardo}\ \bibnamefont {Mazzarello}}, \bibinfo {author}
  {\bibfnamefont {Stefano}\ \bibnamefont {Paolini}}, \bibinfo {author}
  {\bibfnamefont {Alfredo}\ \bibnamefont {Pasquarello}}, \bibinfo {author}
  {\bibfnamefont {Lorenzo}\ \bibnamefont {Paulatto}}, \bibinfo {author}
  {\bibfnamefont {Carlo}\ \bibnamefont {Sbraccia}}, \bibinfo {author}
  {\bibfnamefont {Sandro}\ \bibnamefont {Scandolo}}, \bibinfo {author}
  {\bibfnamefont {Gabriele}\ \bibnamefont {Sclauzero}}, \bibinfo {author}
  {\bibfnamefont {Ari~P.}\ \bibnamefont {Seitsonen}}, \bibinfo {author}
  {\bibfnamefont {Alexander}\ \bibnamefont {Smogunov}}, \bibinfo {author}
  {\bibfnamefont {Paolo}\ \bibnamefont {Umari}}, \ and\ \bibinfo {author}
  {\bibfnamefont {Renata~M.}\ \bibnamefont {Wentzcovitch}},\ }\bibfield
  {title} {\enquote {\bibinfo {title} {{QUANTUM} {ESPRESSO}: a modular and
  open-source software project for quantum simulations of materials},}\ }\href
  {\doibase 10.1088/0953-8984/21/39/395502} {\bibfield  {journal} {\bibinfo
  {journal} {J. Phys.: Condens. Matter}\ }\textbf {\bibinfo {volume} {21}},\
  \bibinfo {pages} {395502} (\bibinfo {year} {2009})}\BibitemShut {NoStop}%
\bibitem [{\citenamefont {Georges}\ \emph {et~al.}(1996)\citenamefont
  {Georges}, \citenamefont {Kotliar}, \citenamefont {Krauth},\ and\
  \citenamefont {Rozenberg}}]{Georges96}%
  \BibitemOpen
  \bibfield  {author} {\bibinfo {author} {\bibfnamefont {Antoine}\ \bibnamefont
  {Georges}}, \bibinfo {author} {\bibfnamefont {Gabriel}\ \bibnamefont
  {Kotliar}}, \bibinfo {author} {\bibfnamefont {Werner}\ \bibnamefont
  {Krauth}}, \ and\ \bibinfo {author} {\bibfnamefont {Marcelo~J.}\ \bibnamefont
  {Rozenberg}},\ }\bibfield  {title} {\enquote {\bibinfo {title} {Dynamical
  mean-field theory of strongly correlated fermion systems and the limit of
  infinite dimensions},}\ }\href {\doibase 10.1103/RevModPhys.68.13} {\bibfield
   {journal} {\bibinfo  {journal} {Rev. Mod. Phys.}\ }\textbf {\bibinfo
  {volume} {68}},\ \bibinfo {pages} {13} (\bibinfo {year} {1996})}\BibitemShut
  {NoStop}%
\bibitem [{\citenamefont {van Loon}\ \emph
  {et~al.}(2014{\natexlab{b}})\citenamefont {van Loon}, \citenamefont
  {Lichtenstein}, \citenamefont {Katsnelson}, \citenamefont {Parcollet},\ and\
  \citenamefont {Hafermann}}]{vanloon_beyond_2014}%
  \BibitemOpen
  \bibfield  {author} {\bibinfo {author} {\bibfnamefont {Erik G. C.~P.}\
  \bibnamefont {van Loon}}, \bibinfo {author} {\bibfnamefont {Alexander~I.}\
  \bibnamefont {Lichtenstein}}, \bibinfo {author} {\bibfnamefont {Mikhail~I.}\
  \bibnamefont {Katsnelson}}, \bibinfo {author} {\bibfnamefont {Olivier}\
  \bibnamefont {Parcollet}}, \ and\ \bibinfo {author} {\bibfnamefont {Hartmut}\
  \bibnamefont {Hafermann}},\ }\bibfield  {title} {\enquote {\bibinfo {title}
  {Beyond extended dynamical mean-field theory: Dual boson approach to the
  two-dimensional extended {Hubbard} model},}\ }\href {\doibase
  10.1103/PhysRevB.90.235135} {\bibfield  {journal} {\bibinfo  {journal} {Phys.
  Rev. B}\ }\textbf {\bibinfo {volume} {90}},\ \bibinfo {pages} {235135}
  (\bibinfo {year} {2014}{\natexlab{b}})}\BibitemShut {NoStop}%
\bibitem [{\citenamefont {Hafermann}\ \emph {et~al.}(2013)\citenamefont
  {Hafermann}, \citenamefont {Werner},\ and\ \citenamefont
  {Gull}}]{Hafermann13}%
  \BibitemOpen
  \bibfield  {author} {\bibinfo {author} {\bibfnamefont {Hartmut}\ \bibnamefont
  {Hafermann}}, \bibinfo {author} {\bibfnamefont {Philipp}\ \bibnamefont
  {Werner}}, \ and\ \bibinfo {author} {\bibfnamefont {Emanuel}\ \bibnamefont
  {Gull}},\ }\bibfield  {title} {\enquote {\bibinfo {title} {Efficient
  implementation of the continuous-time hybridization expansion quantum
  impurity solver},}\ }\href {\doibase
  http://dx.doi.org/10.1016/j.cpc.2012.12.013} {\bibfield  {journal} {\bibinfo
  {journal} {Computer Physics Communications}\ }\textbf {\bibinfo {volume}
  {184}},\ \bibinfo {pages} {1280 -- 1286} (\bibinfo {year}
  {2013})}\BibitemShut {NoStop}%
\bibitem [{\citenamefont {Hafermann}(2014)}]{Hafermann14}%
  \BibitemOpen
  \bibfield  {author} {\bibinfo {author} {\bibfnamefont {Hartmut}\ \bibnamefont
  {Hafermann}},\ }\bibfield  {title} {\enquote {\bibinfo {title} {Self-energy
  and vertex functions from hybridization-expansion continuous-time quantum
  {Monte Carlo} for impurity models with retarded interaction},}\ }\href
  {\doibase 10.1103/PhysRevB.89.235128} {\bibfield  {journal} {\bibinfo
  {journal} {Phys. Rev. B}\ }\textbf {\bibinfo {volume} {89}},\ \bibinfo
  {pages} {235128} (\bibinfo {year} {2014})}\BibitemShut {NoStop}%
\bibitem [{\citenamefont {Bauer}\ \emph {et~al.}(2011)\citenamefont {Bauer},
  \citenamefont {Carr}, \citenamefont {Evertz}, \citenamefont {Feiguin},
  \citenamefont {Freire}, \citenamefont {Fuchs}, \citenamefont {Gamper},
  \citenamefont {Gukelberger}, \citenamefont {Gull}, \citenamefont {Guertler},
  \citenamefont {Hehn}, \citenamefont {Igarashi}, \citenamefont {Isakov},
  \citenamefont {Koop}, \citenamefont {Ma}, \citenamefont {Mates},
  \citenamefont {Matsuo}, \citenamefont {Parcollet}, \citenamefont
  {Pawłowski}, \citenamefont {Picon}, \citenamefont {Pollet}, \citenamefont
  {Santos}, \citenamefont {Scarola}, \citenamefont {Schollwöck}, \citenamefont
  {Silva}, \citenamefont {Surer}, \citenamefont {Todo}, \citenamefont {Trebst},
  \citenamefont {Troyer}, \citenamefont {Wall}, \citenamefont {Werner},\ and\
  \citenamefont {Wessel}}]{ALPS2}%
  \BibitemOpen
  \bibfield  {author} {\bibinfo {author} {\bibfnamefont {B}~\bibnamefont
  {Bauer}}, \bibinfo {author} {\bibfnamefont {L~D}\ \bibnamefont {Carr}},
  \bibinfo {author} {\bibfnamefont {H~G}\ \bibnamefont {Evertz}}, \bibinfo
  {author} {\bibfnamefont {A}~\bibnamefont {Feiguin}}, \bibinfo {author}
  {\bibfnamefont {J}~\bibnamefont {Freire}}, \bibinfo {author} {\bibfnamefont
  {S}~\bibnamefont {Fuchs}}, \bibinfo {author} {\bibfnamefont {L}~\bibnamefont
  {Gamper}}, \bibinfo {author} {\bibfnamefont {J}~\bibnamefont {Gukelberger}},
  \bibinfo {author} {\bibfnamefont {E}~\bibnamefont {Gull}}, \bibinfo {author}
  {\bibfnamefont {S}~\bibnamefont {Guertler}}, \bibinfo {author} {\bibfnamefont
  {A}~\bibnamefont {Hehn}}, \bibinfo {author} {\bibfnamefont {R}~\bibnamefont
  {Igarashi}}, \bibinfo {author} {\bibfnamefont {S~V}\ \bibnamefont {Isakov}},
  \bibinfo {author} {\bibfnamefont {D}~\bibnamefont {Koop}}, \bibinfo {author}
  {\bibfnamefont {P~N}\ \bibnamefont {Ma}}, \bibinfo {author} {\bibfnamefont
  {P}~\bibnamefont {Mates}}, \bibinfo {author} {\bibfnamefont {H}~\bibnamefont
  {Matsuo}}, \bibinfo {author} {\bibfnamefont {O}~\bibnamefont {Parcollet}},
  \bibinfo {author} {\bibfnamefont {G}~\bibnamefont {Pawłowski}}, \bibinfo
  {author} {\bibfnamefont {J~D}\ \bibnamefont {Picon}}, \bibinfo {author}
  {\bibfnamefont {L}~\bibnamefont {Pollet}}, \bibinfo {author} {\bibfnamefont
  {E}~\bibnamefont {Santos}}, \bibinfo {author} {\bibfnamefont {V~W}\
  \bibnamefont {Scarola}}, \bibinfo {author} {\bibfnamefont {U}~\bibnamefont
  {Schollwöck}}, \bibinfo {author} {\bibfnamefont {C}~\bibnamefont {Silva}},
  \bibinfo {author} {\bibfnamefont {B}~\bibnamefont {Surer}}, \bibinfo {author}
  {\bibfnamefont {S}~\bibnamefont {Todo}}, \bibinfo {author} {\bibfnamefont
  {S}~\bibnamefont {Trebst}}, \bibinfo {author} {\bibfnamefont {M}~\bibnamefont
  {Troyer}}, \bibinfo {author} {\bibfnamefont {M~L}\ \bibnamefont {Wall}},
  \bibinfo {author} {\bibfnamefont {P}~\bibnamefont {Werner}}, \ and\ \bibinfo
  {author} {\bibfnamefont {S}~\bibnamefont {Wessel}},\ }\bibfield  {title}
  {\enquote {\bibinfo {title} {The {ALPS} project release 2.0: open source
  software for strongly correlated systems},}\ }\href {\doibase
  10.1088/1742-5468/2011/05/P05001} {\bibfield  {journal} {\bibinfo  {journal}
  {Journal of Statistical Mechanics: Theory and Experiment}\ }\textbf {\bibinfo
  {volume} {2011}},\ \bibinfo {pages} {P05001} (\bibinfo {year}
  {2011})}\BibitemShut {NoStop}%
\bibitem [{\citenamefont {Friedrich}\ \emph {et~al.}(2009)\citenamefont
  {Friedrich}, \citenamefont {Schindlmayr},\ and\ \citenamefont
  {Bl{\"u}gel}}]{friedrich_efficient_2009}%
  \BibitemOpen
  \bibfield  {author} {\bibinfo {author} {\bibfnamefont {Christoph}\
  \bibnamefont {Friedrich}}, \bibinfo {author} {\bibfnamefont {Arno}\
  \bibnamefont {Schindlmayr}}, \ and\ \bibinfo {author} {\bibfnamefont
  {Stefan}\ \bibnamefont {Bl{\"u}gel}},\ }\bibfield  {title} {\enquote
  {\bibinfo {title} {Efficient calculation of the {Coulomb} matrix and its
  expansion around within the {FLAPW} method},}\ }\href {\doibase
  10.1016/j.cpc.2008.10.009} {\bibfield  {journal} {\bibinfo  {journal} {Comp.
  Phys. Comm.}\ }\textbf {\bibinfo {volume} {180}},\ \bibinfo {pages}
  {347--359} (\bibinfo {year} {2009})}\BibitemShut {NoStop}%
\bibitem [{\citenamefont {Friedrich}\ \emph {et~al.}(2010)\citenamefont
  {Friedrich}, \citenamefont {Bl{\"u}gel},\ and\ \citenamefont
  {Schindlmayr}}]{friedrich_efficient_2010}%
  \BibitemOpen
  \bibfield  {author} {\bibinfo {author} {\bibfnamefont {Christoph}\
  \bibnamefont {Friedrich}}, \bibinfo {author} {\bibfnamefont {Stefan}\
  \bibnamefont {Bl{\"u}gel}}, \ and\ \bibinfo {author} {\bibfnamefont {Arno}\
  \bibnamefont {Schindlmayr}},\ }\bibfield  {title} {\enquote {\bibinfo {title}
  {Efficient implementation of the \${GW}\$ approximation within the
  all-electron {FLAPW} method},}\ }\href {\doibase 10.1103/PhysRevB.81.125102}
  {\bibfield  {journal} {\bibinfo  {journal} {Phys. Rev. B}\ }\textbf {\bibinfo
  {volume} {81}},\ \bibinfo {pages} {125102} (\bibinfo {year}
  {2010})}\BibitemShut {NoStop}%
\bibitem [{\citenamefont {Groenewald}\ \emph {et~al.}(2016)\citenamefont
  {Groenewald}, \citenamefont {R{\"o}sner}, \citenamefont {Sch{\"o}nhoff},
  \citenamefont {Haas},\ and\ \citenamefont
  {Wehling}}]{groenewald_valley_2016}%
  \BibitemOpen
  \bibfield  {author} {\bibinfo {author} {\bibfnamefont {R.~E.}\ \bibnamefont
  {Groenewald}}, \bibinfo {author} {\bibfnamefont {M.}~\bibnamefont
  {R{\"o}sner}}, \bibinfo {author} {\bibfnamefont {G.}~\bibnamefont
  {Sch{\"o}nhoff}}, \bibinfo {author} {\bibfnamefont {S.}~\bibnamefont {Haas}},
  \ and\ \bibinfo {author} {\bibfnamefont {T.~O.}\ \bibnamefont {Wehling}},\
  }\bibfield  {title} {\enquote {\bibinfo {title} {Valley plasmonics in
  transition metal dichalcogenides},}\ }\href {\doibase
  10.1103/PhysRevB.93.205145} {\bibfield  {journal} {\bibinfo  {journal} {Phys.
  Rev. B}\ }\textbf {\bibinfo {volume} {93}},\ \bibinfo {pages} {205145}
  (\bibinfo {year} {2016})}\BibitemShut {NoStop}%
\bibitem [{\citenamefont {Sch{\"o}nhoff}\ \emph {et~al.}(2016)\citenamefont
  {Sch{\"o}nhoff}, \citenamefont {R{\"o}sner}, \citenamefont {Groenewald},
  \citenamefont {Haas},\ and\ \citenamefont
  {Wehling}}]{schonhoff_interplay_2016}%
  \BibitemOpen
  \bibfield  {author} {\bibinfo {author} {\bibfnamefont {G.}~\bibnamefont
  {Sch{\"o}nhoff}}, \bibinfo {author} {\bibfnamefont {M.}~\bibnamefont
  {R{\"o}sner}}, \bibinfo {author} {\bibfnamefont {R.~E.}\ \bibnamefont
  {Groenewald}}, \bibinfo {author} {\bibfnamefont {S.}~\bibnamefont {Haas}}, \
  and\ \bibinfo {author} {\bibfnamefont {T.~O.}\ \bibnamefont {Wehling}},\
  }\bibfield  {title} {\enquote {\bibinfo {title} {Interplay of screening and
  superconductivity in low-dimensional materials},}\ }\href {\doibase
  10.1103/PhysRevB.94.134504} {\bibfield  {journal} {\bibinfo  {journal} {Phys.
  Rev. B}\ }\textbf {\bibinfo {volume} {94}},\ \bibinfo {pages} {134504}
  (\bibinfo {year} {2016})}\BibitemShut {NoStop}%
\end{thebibliography}%

\end{document}